\documentclass[journal]{IEEEtran}
\usepackage{amsfonts}
\usepackage{float}
\usepackage{xcolor, soul, framed} 
\usepackage{tabularx}
\usepackage{booktabs}
\colorlet{shadecolor}{yellow}
\usepackage[pdftex]{graphicx}
\usepackage{subcaption}
\graphicspath{{../pdf/}{../jpeg/}}
\DeclareGraphicsExtensions{.pdf,.jpeg,.png}
\usepackage[cmex10]{amsmath}
\usepackage{mathabx}
\usepackage{array}
\usepackage{mdwmath}
\usepackage{mdwtab}
\usepackage{cite}
\usepackage{eqparbox}
\usepackage{url}
\usepackage{multirow}
\usepackage{algorithm}       
\usepackage{algpseudocode}  
\usepackage{braket}
\usepackage{authblk}

\begin{document}
    \bstctlcite{IEEEexample:BSTcontrol}
    \title{Shot and Architecture Adaptive Subspace Variational Quantum Eigensolver for Microwave Simulation}
    \author[1,2]{Zhixiu Han}
    \author[3]{Fanxu Meng}
    \author[1,2,5]{Weidong Li}
    \author[1,2,5,6]{Xutao Yu}
    \author[1,4,5,6]{Zaichen Zhang}

    \affil[1]{School of Information Science and Engineering, Southeast University, Nanjing 211189, Jiangsu, China}
    \affil[2]{State Key Lab of Millimeter Waves, Southeast University, Nanjing 211189, Jiangsu, China}
    \affil[3]{College of Artificial Intelligence, Nanjing Tech University, Nanjing 211800, Jiangsu, China}
    \affil[4]{National Mobile Communications Research Laboratory, Southeast University, Nanjing 211189, Jiangsu, China}
    \affil[5]{Frontiers Science Center for Mobile Information Communication and Security, Nanjing 211111, Jiangsu, China}
    \affil[6]{Quantum Information Center of Southeast University, Nanjing 211189, Jiangsu, China}
    
    


    \markboth{IEEE TRANSACTIONS ON MICROWAVE THEORY AND TECHNIQUES, VOL.~60, NO.~12, DECEMBER~2012 }{Roberg \MakeLowercase{\textit{et al.}}: High-Efficiency Diode and Transistor Rectifiers}

    \maketitle

    \begin{abstract}
            Quantum computing offers a promising paradigm for electromagnetic eigenmode analysis, enabling compact representations of complex field interactions and potential exponential speedup over classical numerical solvers. Recent efforts have applied variational quantum eigensolver (VQE) based methods to compute waveguide modes, demonstrating the feasibility of simulating TE and TM field distributions on noisy intermediate-scale quantum (NISQ) hardware. However, these studies typically employ manually designed, fixed-depth parameterized quantum circuits and uniform measurement-shot strategies, resulting in excessive quantum resource consumption, limited circuit expressivity, and reduced robustness under realistic noise conditions. To address these limitations, we propose an architecture and shot adaptive subspace variational quantum eigensolver for efficient microwave waveguide eigenmode simulation on NISQ devices. 
            The proposed framework integrates a reinforcement learning (RL) based circuit design strategy and an adaptive shot allocation mechanism to jointly reduce quantum resource overhead. Specifically, the RL agent autonomously explores the quantum circuit space to generate hardware-efficient parameterized quantum circuits, while the adaptive measurement scheme allocates sampling resources according to Hamiltonian term weights. 
            Numerical experiments on three- and five-qubit systems demonstrate that the proposed framework achieves accurate estimation of TE and TM mode eigenvalues, with a minimum absolute error down to $10^{-8}$ and reconstructed field distributions under noiseless conditions in excellent agreement with classical electromagnetic solutions. Moreover, robustness tests under realistic noise models confirm the advantages of our algorithm, namely more than a 20-fold speedup in convergence, a reduction of the gate count by up to 45 gates, and consistently lower estimation errors across different noise levels.
            These results establish the proposed framework as a resource-efficient and noise-resistant quantum algorithm for electromagnetic eigenmode analysis in microwave engineering.
    \end{abstract}

    \begin{IEEEkeywords}
   electromagnetic eigenmode analysis, subspace variational quantum eigensolver, reinforcement learning, adaptive shot allocation
    \end{IEEEkeywords}

    %
    \IEEEpeerreviewmaketitle


    \section{Introduction}

\IEEEPARstart{Q}{uantum} computing, empowered by entanglement and superposition, is one of the major transformative technologies. It presents an entirely new computational
paradigm that has the potential to achieve an exponential or polynomial advantage for classically intractable
problems encompassing machine learning \cite{biamonte2017quantum,dunjko2018machine,benedetti2019parameterized,havlivcek2019supervised}, molecular dynamics \cite{mcardle2020quantum,cao2019quantum,bauer2020quantum,ollitrault2021molecular,gaidai2022molecular,mazzola2024quantum}, and combinatorial optimization \cite{farhi2014quantum,blekos2024review,perez2024variational,grange2024introduction,schwagerl2024benchmarking}, etc. With the advent of the noisy intermediate-scale quantum (NISQ) era, where quantum hardware is characterized by limited qubit counts, inherent system noise, significant quantum gate errors, and constrained qubit topology, the community has dedicated itself to exploring quantum algorithms tailored to NISQ devices\cite{preskill2018quantum,bharti2022noisy,chen2023complexity,barligea2025scalability}.
Among these efforts, the variational quantum eigensolver (VQE),
revolving around the use of parameterized quantum circuits (PQCs), emerge as the foremost proposal and has exhibited its superiority in solving tasks across various
domains \cite{yuan2019theory,cerezo2021variational,tilly2022variational}.

The computation of electric and magnetic eigenmodes in waveguides of arbitrary cross section constitutes a long-standing and computationally intensive problem, necessitating the application of sophisticated numerical formulations to manage the associated complexity. The waveguide geometry, dimensional parameters, and material composition govern the modal characteristics. Of particular interest is the theoretical prospect of recasting the conventional numerical solution of the wave equation into a more compact and tractable computational framework, which in principle may be amenable to implementation within a quantum computing paradigm\cite{na2020quantum,jin2024quantum,nguyen2024solving,koukoutsis2023quantum}. 

Recently, by exploiting the VQE framework, Ewe et al. \cite{ewe2022}  pioneered the modeling of TE and TM wave propagation on NISQ devices.
In their approach, the Hamiltonian is decomposed into a linear combination of tensor products of Pauli operators and a cyclic-shift operator. The eigenmodes of the waveguide can then be obtained by evaluating the ground and excited states of this Hamiltonian. Subsequently, Colella et al. \cite{colella2023shot} demonstrated a three-qubit shot-based VQE simulation of waveguide modes that explicitly accounts for measurement effects on quantum hardware. However, these schemes determine the ground and excited states by optimizing a complex nonconvex cost function. This process not only necessitates explicit enforcement of orthogonality between the ground and excited states but also relies on computationally intensive inner product estimations, typically performed via the swap test, which are indispensable in these approaches. To address the above challenges, Miao et al. \cite{miao2024estimation} proposed a subspace-search VQE based waveguide modes, in which an additional orthogonality constraint in the cost function is not required and the computationally expensive inner product evaluation is avoided, typically demanding ancillary qubits and increased circuit depth. Furthermore, the subspace-search VQE scheme has been extended to cylindrical waveguide simulation, as demonstrated in \cite{colella2025Cylindrical}. Nevertheless, the approaches mentioned above typically rely on manually designed PQCs or hardware efficient ansatz (HEA) structure with fixed architectures. The high gate count, especially the number of CNOT gates, imposes a substantial burden on NISQ hardware due to quantum noise; furthermore, the use of a fixed number of measurement shots results in considerable quantum resources being allocated to cost function and gradient evaluations, significantly compromising computational efficiency.

To address the above challenges, this paper proposes an architecture and shot adaptive subspace-search VQE to solve the eigenmodes of rectangular waveguides.
To be precise, we first propose a novel methodology based on RL to enable automated and efficient design of PQC in subspace-search VQE, in which the circuit design is formulated as a Markov process, and through interaction with a quantum environment, the agent explores the circuit design space, avoiding redundant or overly complex structures that are common in hand-crafted and fixed architecture circuits. Second, an adaptive shot allocation strategy is proposed, which adaptively assigns measurement shots across the cost function terms according to the corresponding coefficient weight,  significantly reducing the total measurement overhead as opposed to the fixed-shot strategy employed. Finally, we demonstrate the
superiority of the framework with numerical experiments
over existing methods in terms of gate counts and accuracy under noiseless and noisy situations. 
The results show significant improvements in gate count, accuracy, and convergence speed: the gate count is reduced by up to 45 gates, the error is reduced to $10^{-8}$, and the convergence is more than 20× faster compared with fixed-architecture approaches.

\begin{figure}
    \centering
    \includegraphics[width=0.9\linewidth]{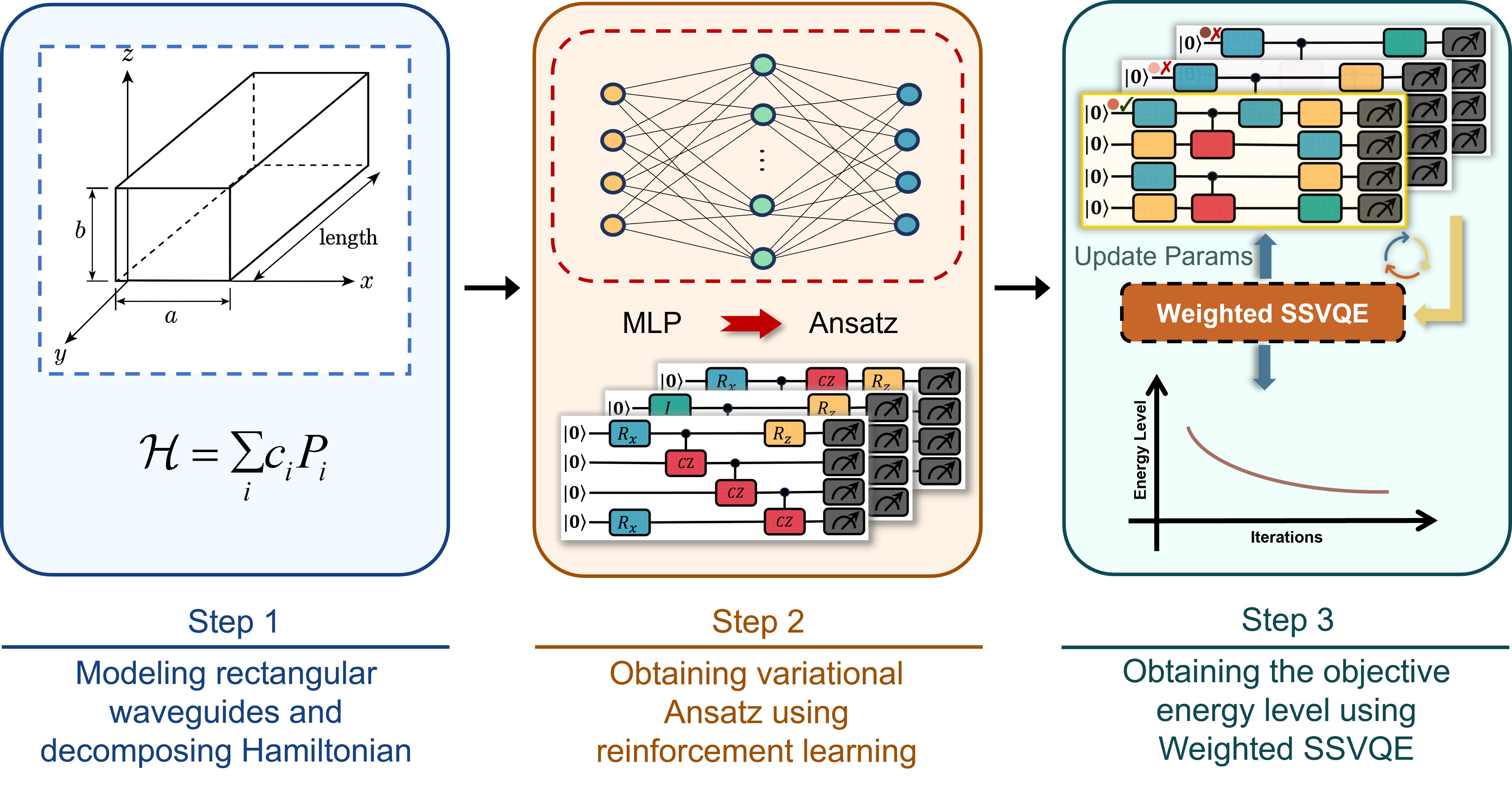}
    \caption{Workflow of the proposed SSVQE-based framework for estimating eigenmodes in rectangular waveguides. }
    \label{fig:placeholder}
\end{figure}
\section{Background}
This section provides a concise overview of the key preliminaries
 underpinning our study. Core notions of quantum computing are first introduced to establish the computational framework. The VQE algorithm is then outlined as a hybrid strategy for ground-state estimation, followed by the SVQE extension that employs subspace expansion for enhanced multi-state accuracy. Finally, key electromagnetic properties of rectangular waveguides are reviewed to provide the physical basis for the subsequent microwave analysis.

\subsection{Quantum Computing}
\label{sec:qc}
Quantum mechanics is formulated in a Hilbert space $\mathcal{H}$, which, for an 
$n$-qubit system, is isomorphic to the complex Euclidean space $\mathbb{C}^{2^n}$. Quantum states are expressed in Dirac notation; a pure state corresponds to a unit-norm column vector $\left| \cdot \right\rangle$, known as a ket. The general $n$-qubit pure state can be written as $\left | \psi  \right \rangle = {\textstyle \sum_{j=1}^{2^n}} \alpha _{j}\left | j  \right \rangle$ where the complex amplitudes satisfy the normalization condition $\sum_{j=1}^{2^n} |\alpha_j|^2 = 1$, and $\left| j \right\rangle$ denotes the computational basis states for $j = 1, 2, \dots, 2^n$.

In quantum computing, quantum gates—represented by unitary matrices—transform one quantum state into another while preserving its norm. Basic gates are typically grouped into single-qubit and two-qubit operations. Common single-qubit gates include the Pauli operators $X, Y, Z, I$ and their associated rotation gates, each of which admits the following unitary-matrix representations:
\begin{align*}
X &= \begin{pmatrix}
0 & 1 \\
1 & 0
\end{pmatrix},
&
R_x(\theta) &= 
\begin{pmatrix}
\cos\left(\frac{\theta}{2}\right) & -i\sin\left(\frac{\theta}{2}\right) \\
-i\sin\left(\frac{\theta}{2}\right) & \cos\left(\frac{\theta}{2}\right)
\end{pmatrix},
\\
Y &= \begin{pmatrix}
0 & -i \\
i & 0
\end{pmatrix},
&
R_y(\theta) &= 
\begin{pmatrix}
\cos\left(\frac{\theta}{2}\right) & -\sin\left(\frac{\theta}{2}\right) \\
\sin\left(\frac{\theta}{2}\right) & \cos\left(\frac{\theta}{2}\right)
\end{pmatrix},
\\
Z &= \begin{pmatrix}
1 & 0 \\
0 & -1
\end{pmatrix},
&
R_z(\phi) &= 
\begin{pmatrix}
e^{-i\frac{\phi}{2}} & 0 \\
0 & e^{i\frac{\phi}{2}}
\end{pmatrix},
I = \begin{pmatrix}
1 & 0 \\
0 & 1
\end{pmatrix}.
\end{align*}

For multi-qubit operations, the Controlled-NOT (CNOT) gate is widely used. It acts on two qubits, flipping the second qubit (target qubit) if the first qubit (control qubit) is in the state $\left| 1 \right\rangle$. The matrix representation of the CNOT gate is:
\[
CNOT = \begin{pmatrix}
1 & 0 & 0 & 0 \\
0 & 1 & 0 & 0 \\
0 & 0 & 0 & 1 \\
0 & 0 & 1 & 0
\end{pmatrix}.
\]

\subsection{Variational Quantum Eigensolver}
\label{VQE}
VQE aims at solving the quantum many-body ground energy problem on NISQ devices. The requisite mathematics of VQE is the Rayleigh-Ritz algorithm\cite{mcclean2016theory}, which minimizes the objective by optimizing the parameters of the wave function:
\begin{equation}
        E\left(\boldsymbol{\vec{\theta}  } \right) = \min_{\boldsymbol{\vec{\theta}  }}\left \langle \psi_{0}    \right | U^{\dagger }\left ( \boldsymbol{\vec{\theta}  } \right )\hat{H}U\left ( \boldsymbol{\vec{\theta}  } \right )\left | \psi _{0}  \right \rangle,  
\end{equation}
where the initial state $\left | \psi _{0}  \right \rangle $ is generally  $\left | 00\dots 0  \right \rangle$, $\hat{H}$ is the system Hamiltonian and $U\left ( \boldsymbol{\vec{\theta}  } \right )$ is the unitary of the employed PQC. In VQE, the objective is calculated by the measurement operation on the quantum device and fed into the classical optimizer to fine-tune the parameters. In particular, the system Hamiltonian can often be expressed as a linear combination of Pauli operators, as follows:
\begin{equation}
\hat{H} = \sum_{i=1}^{l} \alpha_i P_i,
\label{eq:H_pauli}
\end{equation}
where \(\alpha_i\) are real coefficients, and \(P_i\) are Pauli operators (\(I\), \(X\), \(Y\), \(Z\)). This decomposition provides a compact and efficient representation for simulating quantum systems.

\subsection{Subspace Variational Quantum Eigensolver}

Subspace variational quantum eigensolver (SSVQE) extends the conventional VQE framework to compute ground and excited states simultaneously\cite{nakanishi2019subspace,heya2019subspace}. This algorithm initializes a set of mutually orthogonal states and exploits the inherent unitarity of parameterized quantum circuits to maintain their orthogonality throughout the optimization. As a result, the procedure naturally converges to a low-energy subspace encompassing the target states. 
    Specifically, the algorithm first maps the input states into the low-energy subspace of the Hamiltonian through an initial optimization, and then extracts the target excited states via a second optimization. 
    In the weighted SSVQE variant, only a single optimization as shown in Eq.~\eqref{E_w} is required to obtain either the $k$-th excited state or the lowest $k$ excited states. 
    Compared with other VQE variants, weighted SSVQE avoids inner product measurements and auxiliary qubits, which enhances its practicality for NISQ devices.
    \begin{equation}
    E\left( \boldsymbol{\vec{\theta}  } \right) = \min_{\boldsymbol{\vec{\theta}  }}\sum_{j=0}^k w_j \left \langle \psi_j | U^\dagger\left(\boldsymbol{\vec{\theta}  }\right) \hat{H} U\left(\boldsymbol{\vec{\theta} }\right) | \psi_j \right \rangle
    \label{E_w}
    \end{equation}
The workflow of weighted SSVQE can be summarized as follows:
\begin{figure}[htbp]
    \centering
    \includegraphics[width=0.9\linewidth]{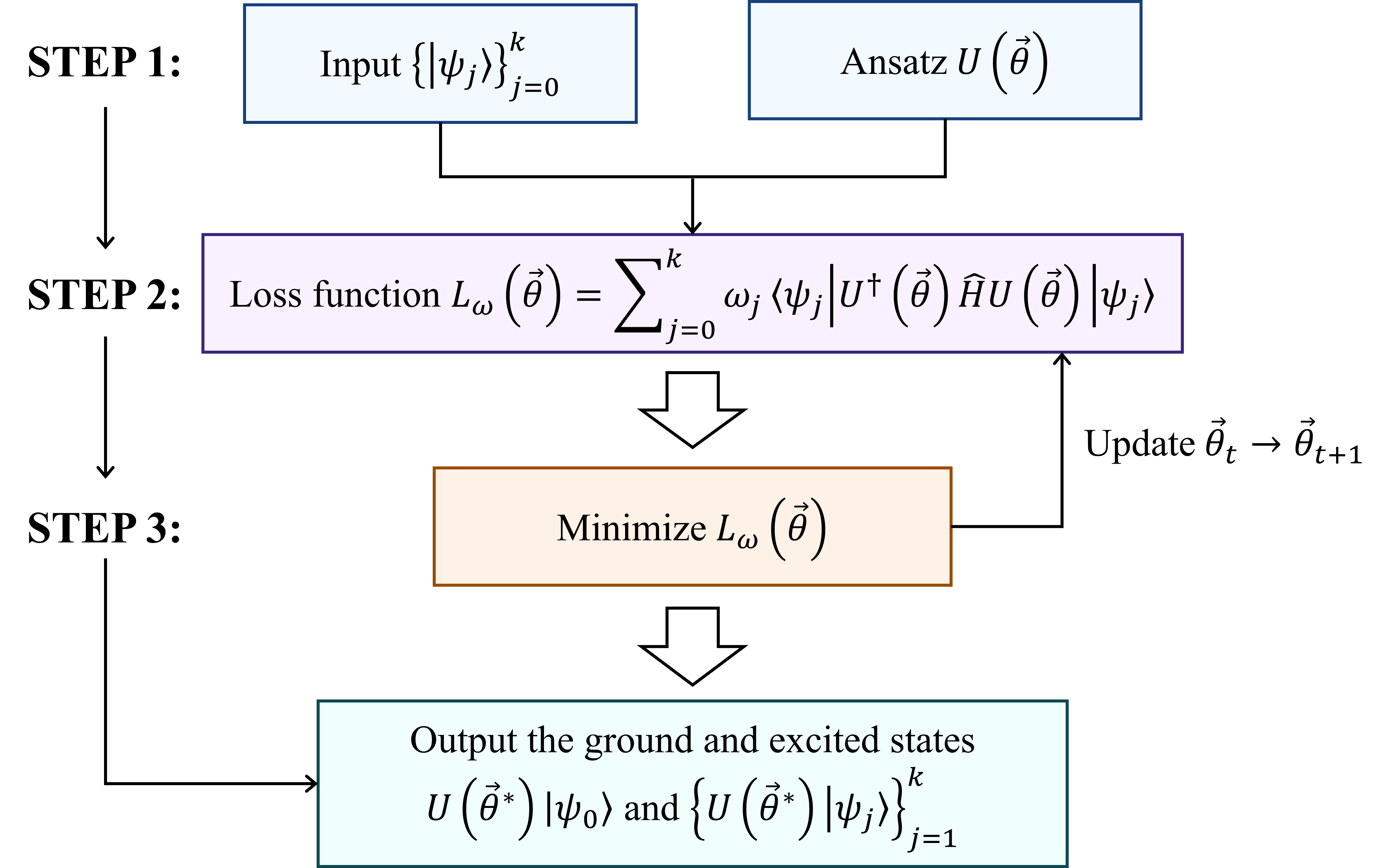}
    \caption{Weighted SSVQE optimization workflow.}
    \label{fig:placeholder}
\end{figure}

\begin{algorithm}[htbp]
\caption{Subspace-search variational quantum eigensolver.}
\begin{algorithmic}[1]
\item Select a set of mutually orthogonal trial states \( \{ |\psi_j\rangle \}_{j=0}^{k} \) and construct a parameterized quantum circuit \( U(\vec{\theta}) \).
\item Initialize $\vec{\theta}$ randomly and update the $\vec{\theta}$ using a classical optimizer to solve \(
    E\left( \boldsymbol{\vec{\theta}  } \right) = \min_{\boldsymbol{\vec{\theta}  }}\sum_{j=0}^k w_j \left \langle \psi_j | U^\dagger\left(\boldsymbol{\vec{\theta}  }\right) \hat{H} U\left(\boldsymbol{\vec{\theta} }\right) | \psi_j \right \rangle,
    \) where \(\{w_j\}\) satisfies \(w_i > w_j\) for \(i < j\).

\item Output the ground and excited states $U(\vec{\theta}^*) \ket{\psi_0}$ and $\{U(\vec{\theta}^*) \ket{\psi_j}\}_{j=1}^k$ where $\vec{\theta}^*$ is the optimal parameters.
\end{algorithmic}
\end{algorithm}

\subsection{Rectangular Waveguides}

For a rectangular metal waveguide filled with a uniform material having dielectric constant $\epsilon_0$ and a magnetic permeability $\mu_0$, the Helmholtz equations can be used to describe the field distribution in a waveguide in the case of TE and TM waves:
        \begin{equation}
        \nabla_{s}^{2}E_{z}= -k_{c}^{2}E_{z}, \text{TM case,}
        \label{tm}
    \end{equation}
    \begin{equation}
        \nabla_{s}^{2}H_{z}= -k_{c}^{2}H_{z}, \text{TE case.}
        \label{te}
    \end{equation}
    where $k_{c}^{2}= k^{2}- k_{z}^{2}$, $k = \sqrt{\omega^{2}\mu_0 \epsilon_0}$
    represents the wavenumber, $\omega$ is the angular frequency, and $k_{z}$ is
    the propagation constant along the $z$-direction. The modal solutions within the metallic waveguide are differentiated by the boundary conditions imposed on their respective longitudinal field components at the structure's walls. The axial electric field, $E_z$, for TM modes is constrained by a homogeneous Dirichlet condition ($E_z = 0$). Meanwhile, the axial magnetic field, $H_z$, for TE modes is subjected to a homogeneous Neumann condition, requiring its normal derivative to be null ($\partial H_z / \partial n = 0$).

Assuming a solution of the form $F_z(x,y) = X(x)Y(y)(F_z\in[E_{z},H_{z}])$, the method of separation of variables decomposes each two-dimensional Helmholtz equation into two one-dimensional ordinary differential equations for the $x$ and $y$ coordinates respectively. Each of these equations takes the standard form:
\begin{equation}
    -\frac{d^2 f(l)}{dl^2} = k_l^2 f(l)
    \label{sv}
\end{equation}
where $f(l)$ represents the separated function ($X(x)$ or $Y(y)$) of the longitudinal field component ($E_z$ or $H_z$), $l$ is the spatial variable ($x$ or $y$), and $k_l^2$ is the corresponding separation constant.

In applying the Finite Difference Method (FDM), the partial derivatives within the Eq.~\eqref{sv} are approximated by finite difference expressions. 
The second-order partial derivatives in the Laplacian operator are approximated using the second-order central difference scheme:

\begin{equation}
    \frac{d^2 f(l)}{dl^2}\bigg|_{i} \approx \frac{f_{i+1} - 2f_i + f_{i-1}}{(\Delta l)^2},
    \label{fdm}
\end{equation}
where $f_i$ represents the value of the function at the central node $i$, while $f_{i+1}$ and $f_{i-1}$ are the values at the adjacent nodes.

By substituting Eq.~\eqref{fdm} into Eq.~\eqref{sv}, the differential equation is transformed into a discrete algebraic form:
\begin{equation}
        Af=k_{l}^{2}f.
        \label{A}
    \end{equation}
This algebraic equation represents a discrete eigenvalue problem, where $A$ is the coefficient matrix derived from the finite difference scheme, $f$ is the eigenvector representing the discrete field values of the mode, and $k_{l}^{2}$ is the corresponding eigenvalue.

\section{Architecture and Shot Adaptive Subspace VQE Algorithm}

In this section, we detail the overall structure of the proposed framework. First, we develop a Hamiltonian decomposition strategy for rectangular waveguide eigenmodes, enabling their representation in a form compatible with SSVQE. Next, we introduce an RL-based quantum circuit design module that automatically constructs resource-efficient circuit. Finally, we incorporate an adaptive-shot measurement scheme to reduce sampling overhead during the parameters optimization process.

    \subsection{Hamiltonian Decomposition For Waveguide Eigenmodes}
    The discretized system matrix $A$ in Eq.~\eqref{A} can be interpreted as the Hamiltonian $H$ matrix of a corresponding multi-qubit system. For a three-qubit system, the $2^{3}
    \times 2^{3}$ Hamiltonian matrix for the TM case is expressed as follows:
    \begin{equation}
        H_{\text{TM}}= \frac{1}{(\Delta l)^2}
        \begin{bmatrix}
            3      & -1     & 0      & \cdots & 0      \\
            -1     & 2      & -1     & \cdots & 0      \\
            \vdots &        & \ddots &        & \ddots \\
            0      & \cdots & -1     & 2      & -1     \\
            0      & \cdots & 0      & -1     & 3
        \end{bmatrix}.
    \end{equation}

    Similarly, in the case of TE mode, the
    Hamiltonian matrix is expressed as follows:

    \begin{equation}
        H_{\text{TE}}= \frac{1}{(\Delta l)^2}
        \begin{bmatrix}
            1      & -1     & 0      & \cdots & 0      \\
            -1     & 2      & -1     & \cdots & 0      \\
            \ddots &        & \ddots &        & \ddots \\
            0      & \cdots & -1     & 2      & -1     \\
            0      & \cdots & 0      & -1     & 1
        \end{bmatrix}.
    \end{equation}

The Hamiltonian matrix $H_{\text{TM}}$ and $H_{\text{TE}}$ can be decomposed into a linear combination of the following Pauli operators:
\begin{equation}
\begin{aligned}
\hat{H}_{\text{TM}} &=  (2.25III - 1.0IIX - 0.5IXX \\
             & - 0.5IYY + 0.25IZZ - 0.25XXX \\
             & + 0.25XYY - 0.25YXY - 0.25YYX \\
             & + 0.25ZIZ + 0.25ZZI)/ (\Delta l)^2
\end{aligned}
\end{equation}

\begin{equation}
\begin{aligned}
\hat{H}_{\text{TE}} &=  (1.75III - 1.0IIX - 0.5IXX \\
             & - 0.5IYY - 0.25IZZ - 0.25XXX \\
             & + 0.25XYY - 0.25YXY - 0.25YYX \\
             & - 0.25ZIZ - 0.25ZZI)/ (\Delta l)^2
\end{aligned}
\end{equation}

In a five-qubit quantum system, the Hamiltonian matrix can also be decomposed. While the increased dimensionality renders the decomposition more complex, it enables a more precise characterization of the field distribution within a rectangular waveguide. The complete Hamiltonian encompasses 47 distinct terms. For the sake of conciseness, only a subset of these terms is provided in this discussion.
\begin{equation}
\begin{aligned}
\hat{H}_{\text{TM}} &=  2.0625 IIIII - 1.0 IIIIX - 0.5 IIIXX \\
             & - 0.5 IIIYY + 0.0625 IIIZZ \\
             & \vdots \\
             & + 0.0625 ZIZZZ + 0.0625 ZZIII + 0.0625 ZZIZZ \\
             & + 0.0625 ZZZIZ + 0.0625 ZZZZI
\label{13}
\end{aligned}
\end{equation}
\begin{equation}
\begin{aligned}
\hat{H}_{\text{TE}} &=  1.9375 IIIII - 1.0 IIIIX - 0.5 IIIXX \\
             & - 0.5 IIIYY - 0.0625 IIIZZ \\
             & \vdots \\
             & + 0.0625 ZIZZZ + 0.0625 ZZIII + 0.0625 ZZIZZ \\
             & + 0.0625 ZZZIZ + 0.0625 ZZZZI
\label{14}
\end{aligned}
\end{equation}

\subsection{RL-based Quantum Circuit Design}
\begin{figure}[htbp]
    \centering
    \includegraphics[width=0.75\linewidth]{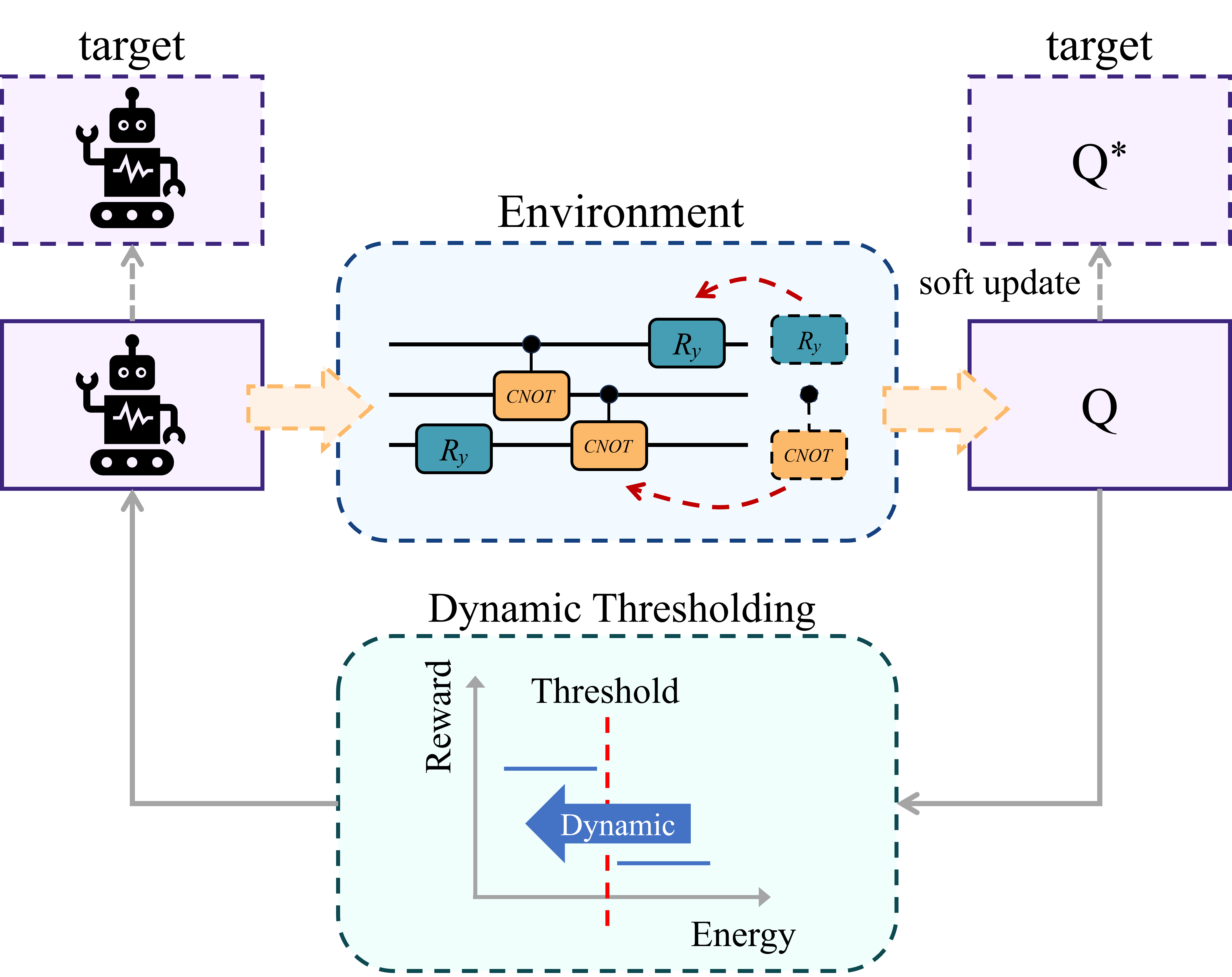}
    \caption{DDQN-based RL framework for automated circuit design.}
    \label{fig:rl}
\end{figure}

In SSVQE, the architecture of parameterized quantum circuit is critical, as it directly determines whether the algorithm can effectively approximate the target ground and excited states. Existing implementations predominantly rely on manually designed PQCs or HEA structures, both of which employ fixed circuit layouts, imposing a substantial burden on NISQ hardware due to limited coherence times and accumulated quantum noise. Therefore, in this work, RL is employed to automatically construct quantum circuit in SSVQE, as schematically illustrated in Fig.~\ref{fig:rl}. RL is capable of efficiently exploring the
quantum circuit space to generate hardware-efficient parameter
ized quantum circuits. 

\begin{figure}[htbp]
    \centering
    \includegraphics[width=0.9\linewidth]{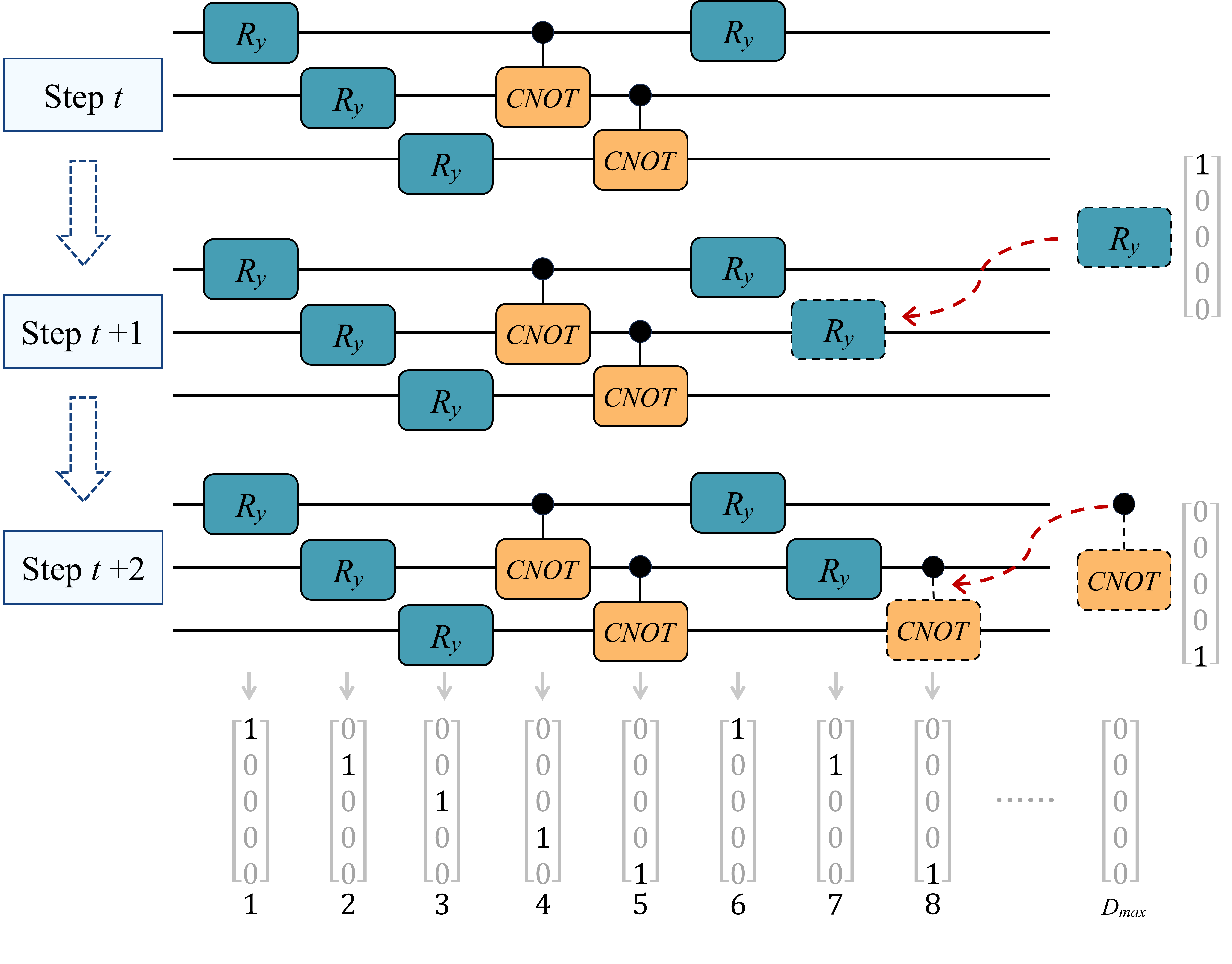}
    \caption{One-hot encoding of the actions and sequential action selection of quantum gates at each time step}
    \label{fig:action}
\end{figure}

To achieve the automated design of a variational quantum circuit, we formulate this problem as a Markov Decision Process (MDP) defined by the tuple $(\mathcal{S}, \mathcal{A}, P, R, \gamma)$. The MDP provides a mathematical framework for sequential decision-making problems, the core of which lies in mapping the construction of the quantum circuit to an agent-environment interaction model. 

The state space $\mathcal{S}$ is designed to encode the quantum circuit architecture. Let $D_{max}$ be the maximum allowed circuit depth and $N_A$ be the total number of available actions. The structural component of the state is represented as a flattened vector derived from a one-hot encoded matrix of shape $D_{max} \times N_A$. 
As illustrated in Fig.~\ref{fig:action}, if the $m$-th action is selected at depth $d$, the component at index $d \times N_A + m$ is set to 1, while all other entries remain 0. 
To provide direct feedback on the circuit's quality, this structural vector is concatenated with the current weighted energy value.

To demonstrate the advantage of our proposed architecture search over fixed HEA, we restrict our action space to the same gate set used in standard HEA.
The discrete action space is constructed by enumerating all possible placements of these gates on the $N$-qubit system. The single-qubit actions are given by $\mathcal{A}_{1} = \{R_Y^i \mid i \in \{0, \dots, N-1\}\}$, and the two-qubit entangling actions are restricted to adjacent pairs, defined as $\mathcal{A}_{2} = \{\text{CNOT}_{i, i+1} \mid i \in \{0, \dots, N-2\}\}$. The full action space is the union $\mathcal{A} = \mathcal{A}_{1} \cup \mathcal{A}_{2}$.

As shown in Fig.~\ref{fig:action}, at each time step $t$, the agent selects an action $a_t$ from the action space $\mathcal{A}$ according to an $\epsilon$-greedy policy. This means the agent usually selects the gate with the highest predicted Q-value to exploit known good strategies, but occasionally selects a random gate with probability $\epsilon$ to explore new circuit structures.

The agent is incentivized to minimize the weighted energy objective $E(\boldsymbol{\vec{\theta}})$ as defined in Eq.~\eqref{E_w}. The immediate reward $r_t$ is designed to facilitate efficient search within the solution space, defined by:
\begin{equation}
    r_t = - E(\boldsymbol{\vec{\theta}}) - \lambda \cdot D_t
\end{equation}
where $D_t$ represents the current circuit depth and $\lambda$ is a penalty factor to encourage compact circuits.

Crucially, to prevent the agent from converging to suboptimal local minima, we implement a dynamic reward strategy based on curriculum learning. As the training progresses, we maintain a success threshold $\xi$ that dynamically decreases based on the agent's best historical performance. A positive bonus is assigned only when the current circuit energy is lower than $\xi$. This time-varying reward mechanism encourages the agent to continuously explore lower-energy subspaces rather than settling for easily accessible but suboptimal solutions.

Given this dynamic reward structure, we employ a Double Deep Q-Network (DDQN) to optimize the policy and stabilize training. The agent updates its policy by minimizing the Mean Squared Error (MSE) between the predicted Q-values and the target Q-values. As implemented in our algorithm, the target Q-value $y_t$ is computed using a separate target network:
\begin{equation}
    y_t = r_t + \gamma \max_{a'} Q(S_{t+1}, a'; \boldsymbol{\omega}^-)
\end{equation}
where $\boldsymbol{\omega}^-$ denotes the parameters of the target network. Crucially, to ensure training stability, the parameters of the target network are not updated directly but slowly track the online network parameters via a soft update mechanism ($\boldsymbol{\omega}^- \leftarrow \tau \boldsymbol{\omega} + (1-\tau)\boldsymbol{\omega}^-$).

To optimize the allocation of measurement resources, we propose an adaptive shot allocation mechanism.
By substituting the Pauli decomposition from Eq.~\eqref{eq:H_pauli} (with coefficients denoted by $\alpha_i$) into the loss function in Eq.~\eqref{E_w}, and defining a combined coefficient $c_{j,i} = w_j \alpha_i$, we can rewrite Eq.~\eqref{E_w} as:
\begin{equation}
E\left( \boldsymbol{\vec{\theta} } \right)
  = \sum_{j=0}^k \sum_{i=1}^{l} c_{j,i} \left \langle \psi_j \middle| U^\dagger\left(\boldsymbol{\vec{\theta} }\right) P_i U\left(\boldsymbol{\vec{\theta} }\right) \middle| \psi_j \right \rangle.
\label{E_expanded}
\end{equation}

We assume that the signs of $c_{j,i}$ are absorbed into the measurement results.
Let $\mathbf{R}$ be a random variable representing the selection of the term index pair $(j, i)$ corresponding to the state $\psi_j$ and Pauli operator $P_i$. The probability distribution of $\mathbf{R}$ is defined as:
\begin{equation}
\mathbf{R} =
\begin{cases}
(\psi_0, P_1) & \text{with probability } p_{0,1} =
      \dfrac{|c_{0,1}|}{\lVert \mathbf{c} \rVert_{\ell_1}}, \\
(\psi_0, P_2) & \text{with probability } p_{0,2} =
      \dfrac{|c_{0,2}|}{\lVert \mathbf{c} \rVert_{\ell_1}}, \\
\vdots & \\
(\psi_j, P_i) & \text{with probability } p_{j,i} =
      \dfrac{|c_{j,i}|}{\lVert \mathbf{c} \rVert_{\ell_1}}, \\
\vdots & \\
(\psi_k, P_l) & \text{with probability } p_{k,l} =
      \dfrac{|c_{k,l}|}{\lVert \mathbf{c} \rVert_{\ell_1}},
\end{cases}
\label{sampling_prob}
\end{equation}
where $\mathbf{c}$ denotes the tensor of all combined coefficients, and $\lVert \mathbf{c} \rVert_{\ell_1} = \sum_{j=0}^k \sum_{i=1}^l |c_{j,i}|$
is the $\ell_1$-norm. The random variable $\mathbf{R}$ implies that each term is selected with probability proportional to the magnitude of its coefficient $c_{j,i}$.

Consequently, we rewrite the loss function in Eq.~\eqref{E_w} using the expectation value over $\mathbf{R}$:
\begin{equation}
E\left( \boldsymbol{\vec{\theta} } \right) =
\mathbb{E}\!\left[
  \lVert \mathbf{c} \rVert_{\ell_1} \cdot
  \left\langle \psi_{j} \middle|
  U^\dagger\left(\boldsymbol{\vec{\theta} }\right) \mathbf{R} \, U\left(\boldsymbol{\vec{\theta} }\right) \middle| \psi_{j}
  \right\rangle
\right],
\label{E_importance}
\end{equation}
where in the expectation term, $\mathbf{R}$ acts as the selected Pauli operator $P_i$ associated with the selected state $\psi_j$.
This formulation ensures that terms contributing more significantly to the total energy are measured more frequently, thereby realizing an adaptive shot strategy that effectively reduces the estimation cost.
\begin{figure*}[t]
    \centering
    \begin{subfigure}[b]{0.24\textwidth}
        \centering
        \includegraphics[width=\textwidth]{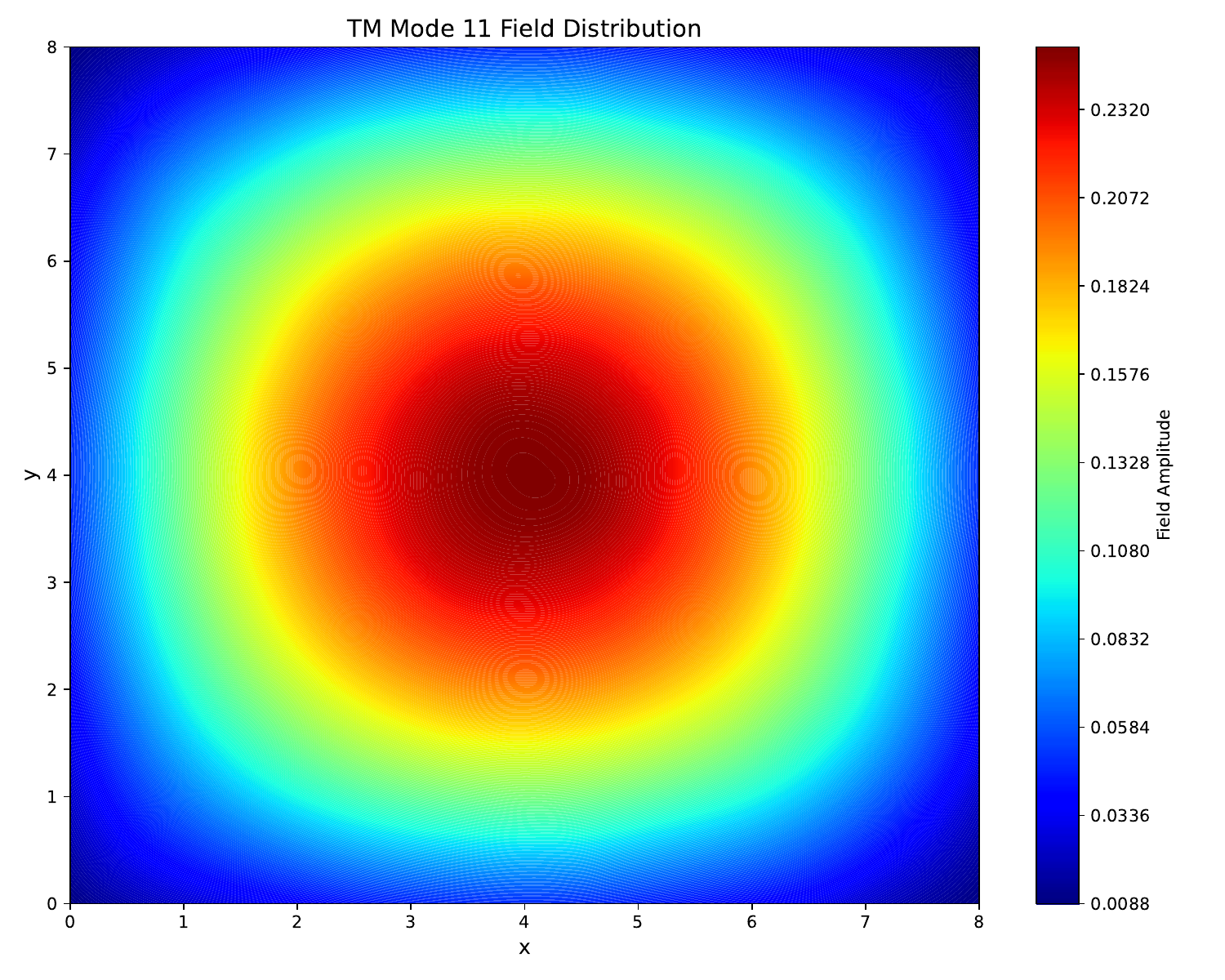} 
    \end{subfigure}
    \begin{subfigure}[b]{0.24\textwidth}
        \centering
        \includegraphics[width=\textwidth]{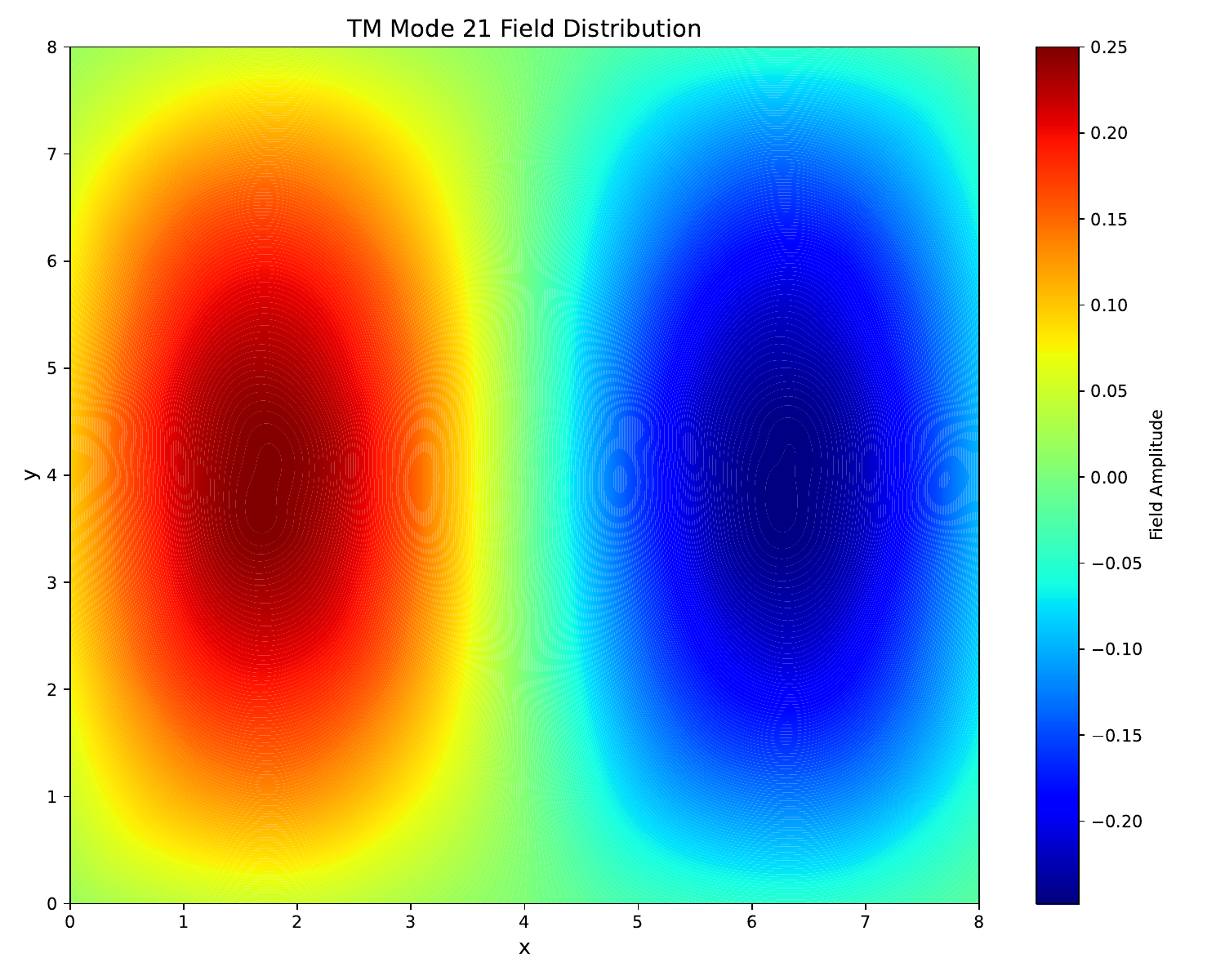} 
    \end{subfigure}
    \begin{subfigure}[b]{0.24\textwidth}
        \centering
        \includegraphics[width=\textwidth]{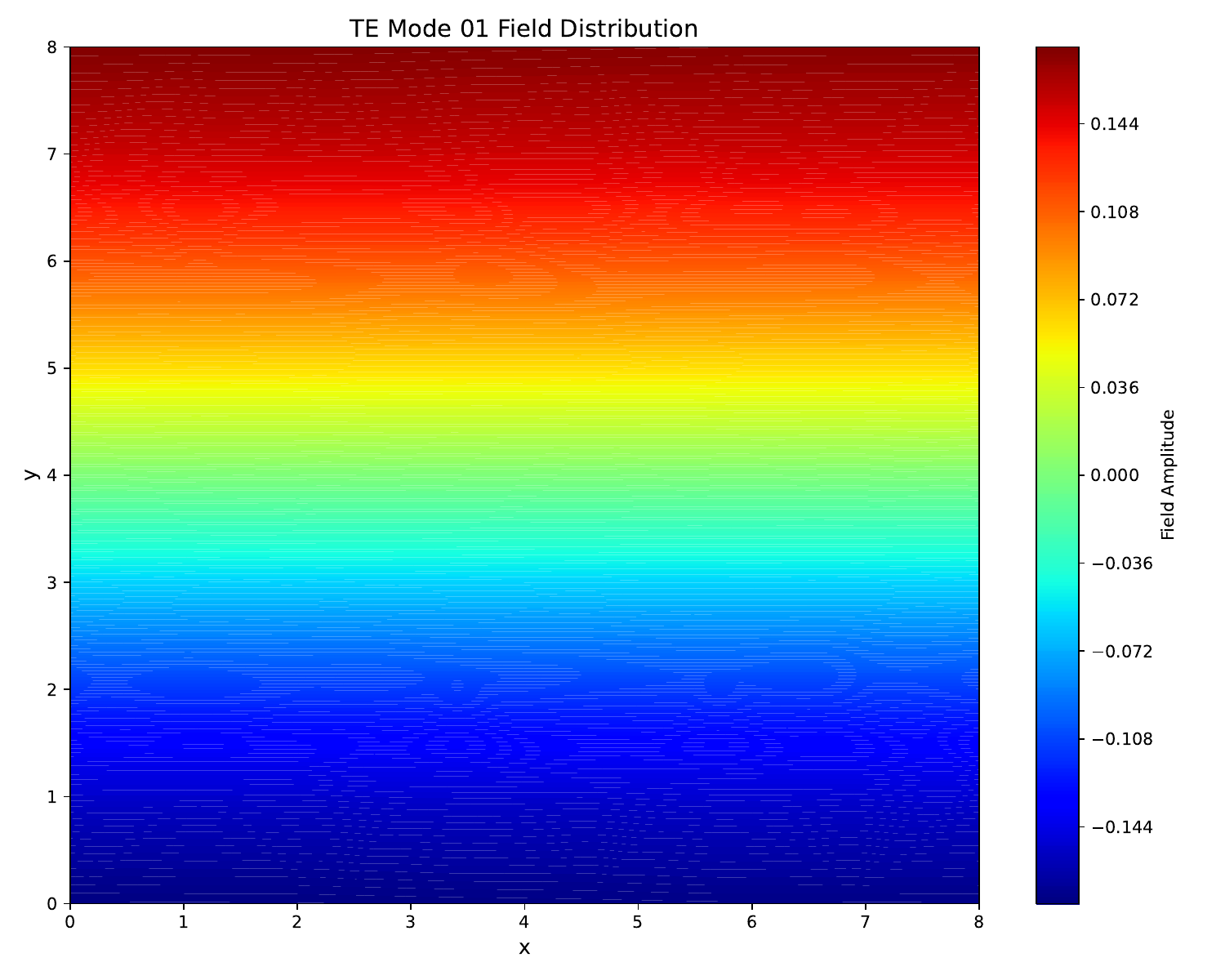}
    \end{subfigure}
    \begin{subfigure}[b]{0.24\textwidth}
        \centering
        \includegraphics[width=\textwidth]{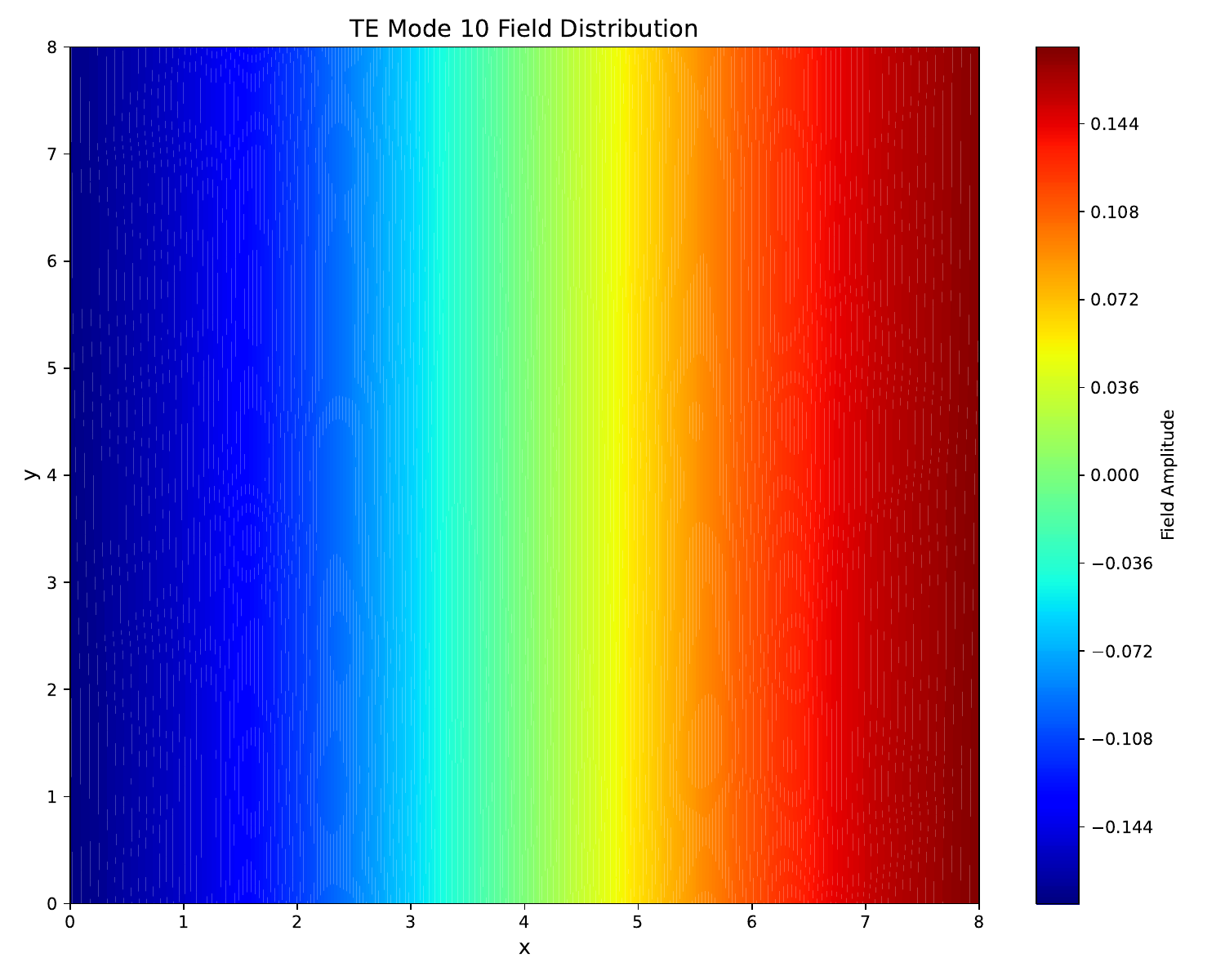} 
    \end{subfigure}
    \caption{Reconstructed field distributions for three-qubit waveguide modes from the obtained eigenstates.}
    \label{fig:field_ideal_3q}
\end{figure*}

\begin{figure*}[t]
    \centering
    \begin{subfigure}[b]{0.24\textwidth}
        \centering
        \includegraphics[width=\textwidth]{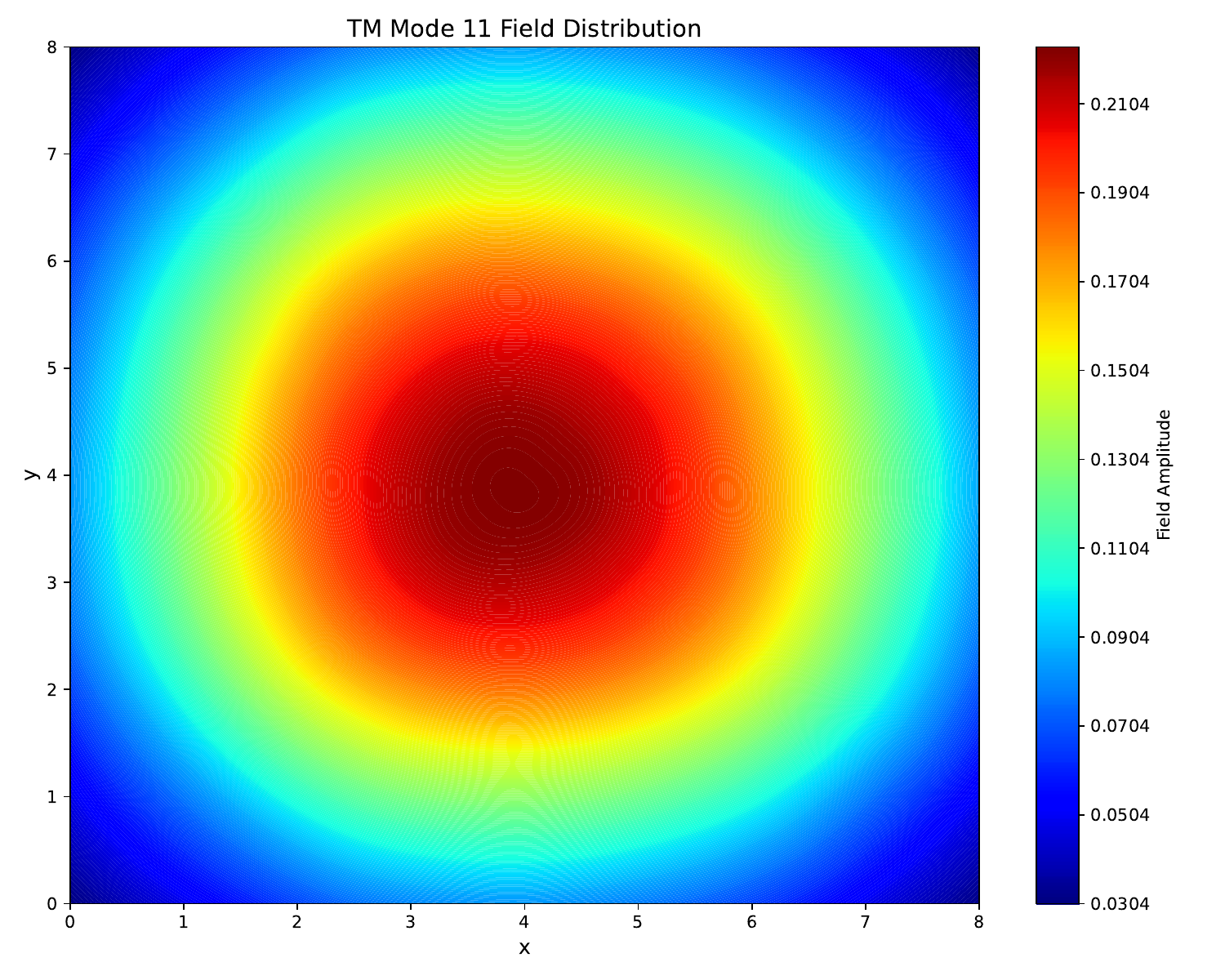} 
    \end{subfigure}
    \begin{subfigure}[b]{0.24\textwidth}
        \centering
        \includegraphics[width=\textwidth]{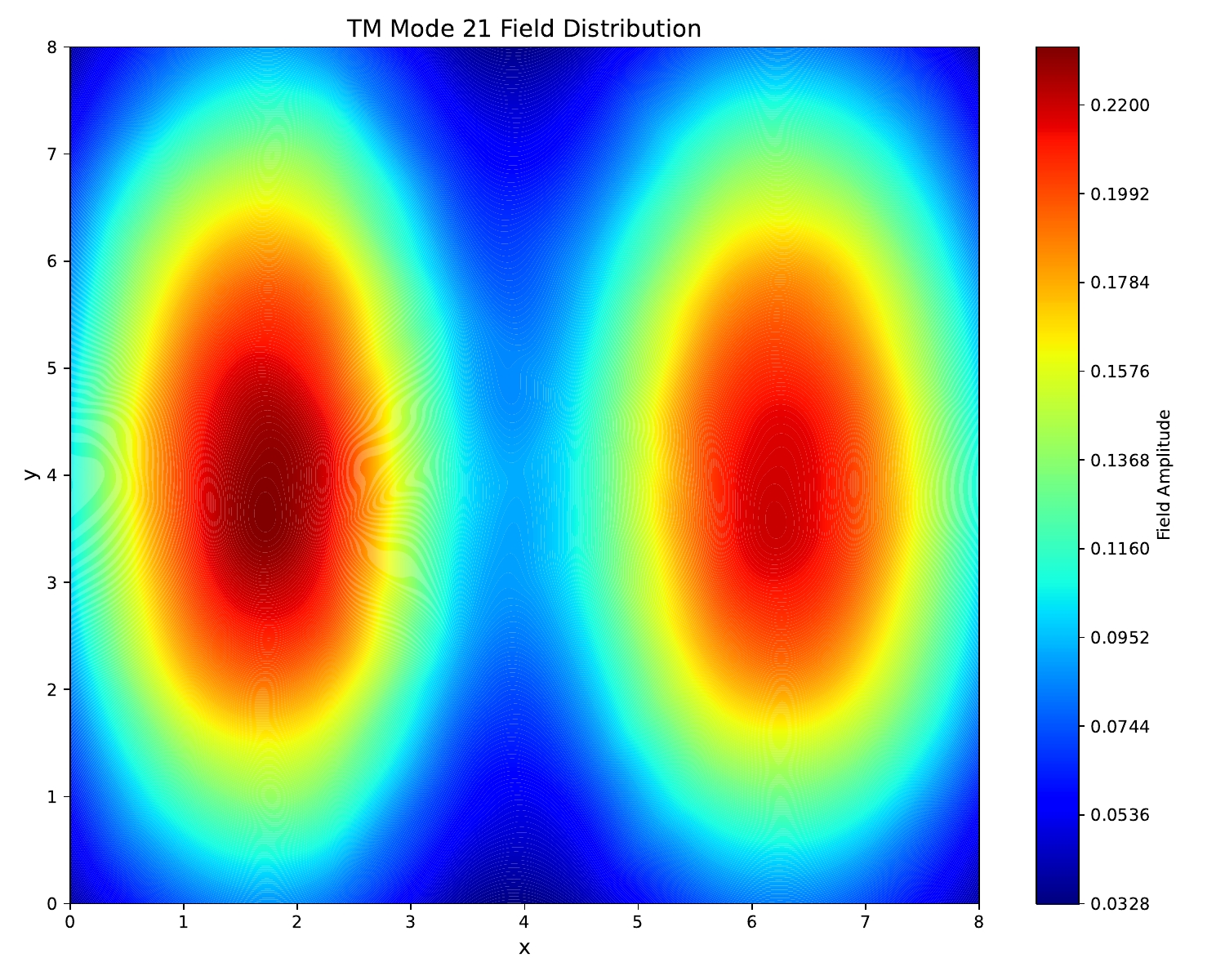} 
    \end{subfigure}
    \begin{subfigure}[b]{0.24\textwidth}
        \centering
        \includegraphics[width=\textwidth]{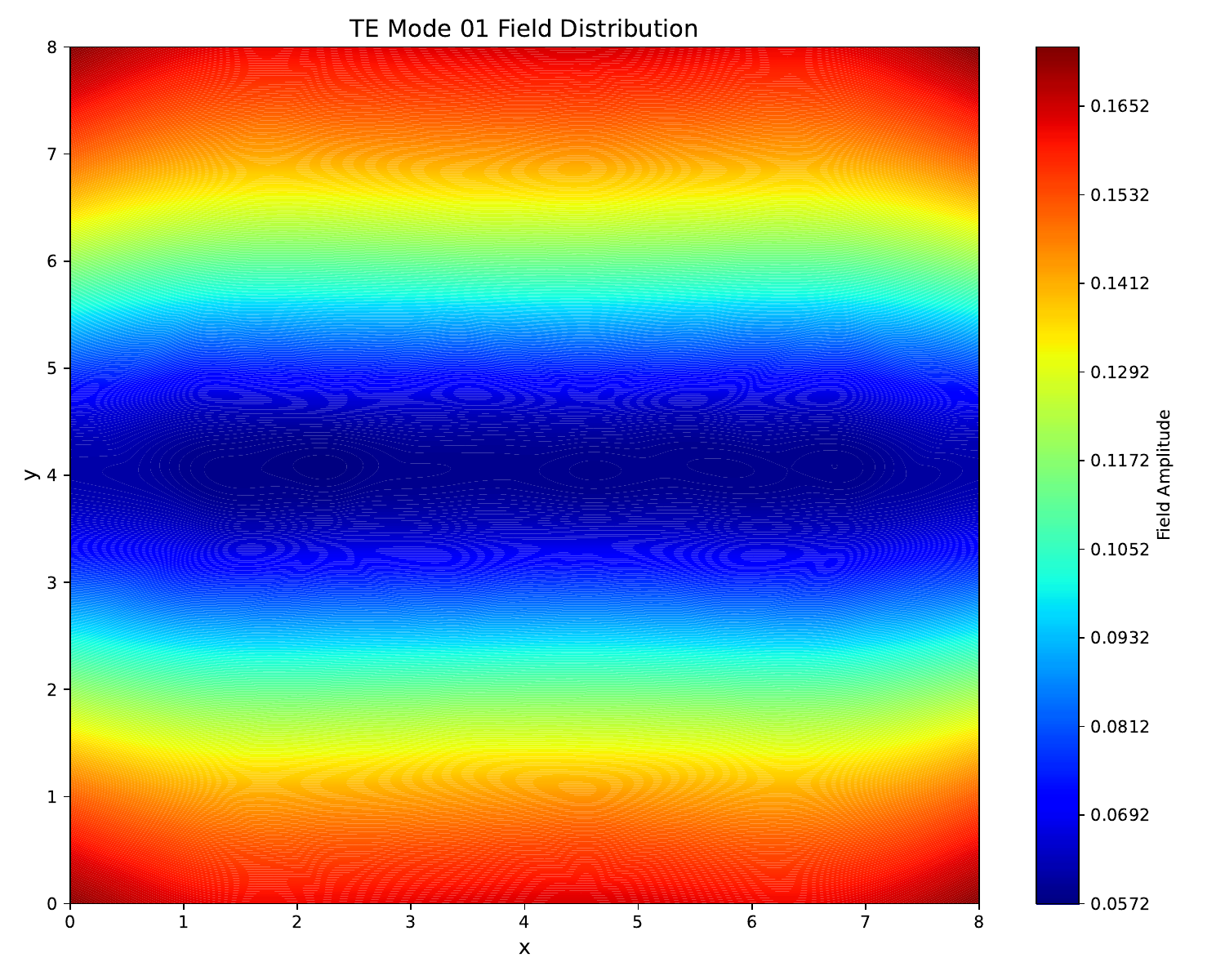}
    \end{subfigure}
    \begin{subfigure}[b]{0.24\textwidth}
        \centering
        \includegraphics[width=\textwidth]{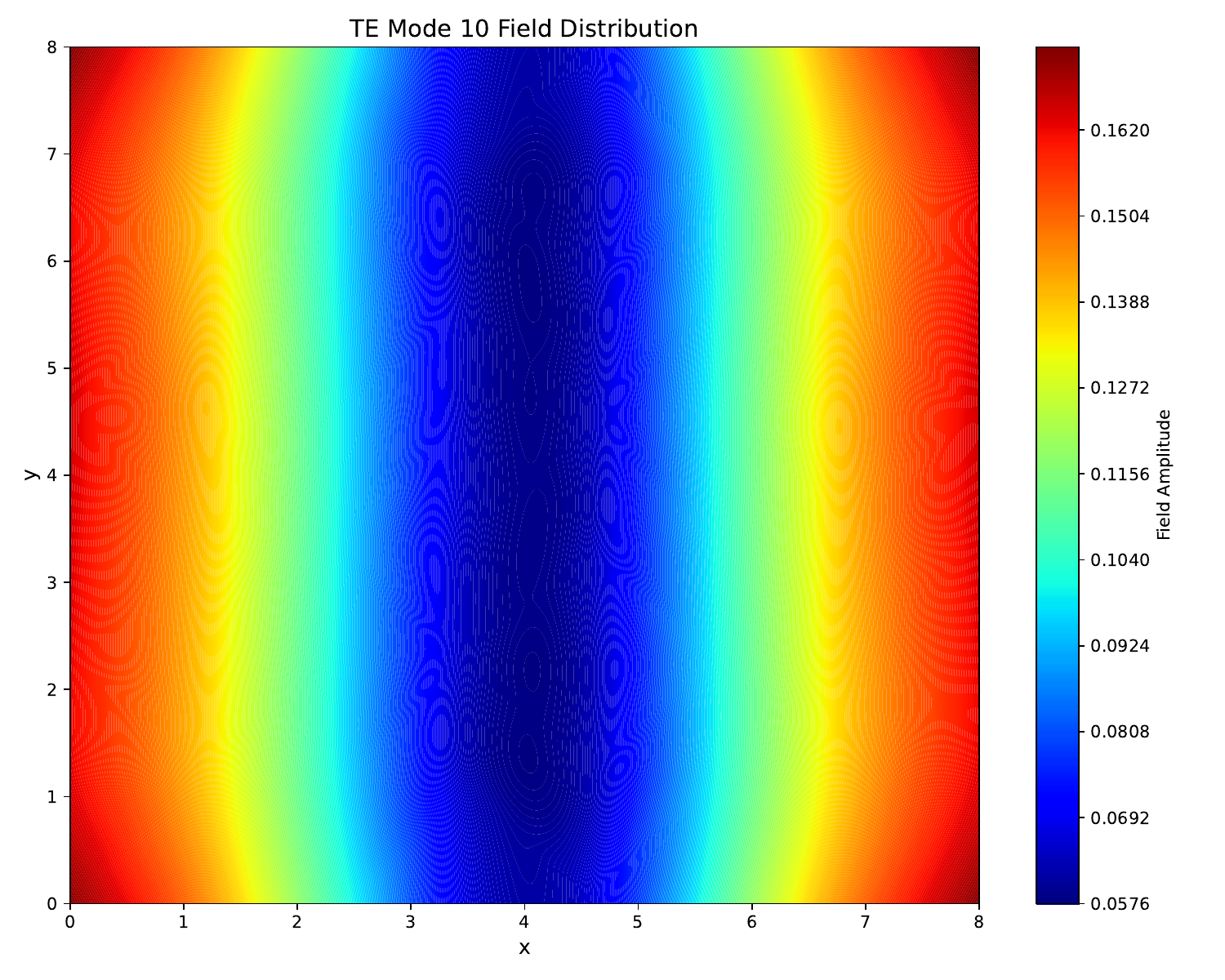} 
    \end{subfigure}
    \caption{Field reconstruction under depolarizing noise for three-qubit systems.}
    \label{fig:field_noise_3q}
\end{figure*}
\begin{figure*}[t]
    \centering
    \begin{subfigure}[b]{0.24\textwidth}
        \centering
        \includegraphics[width=\textwidth]{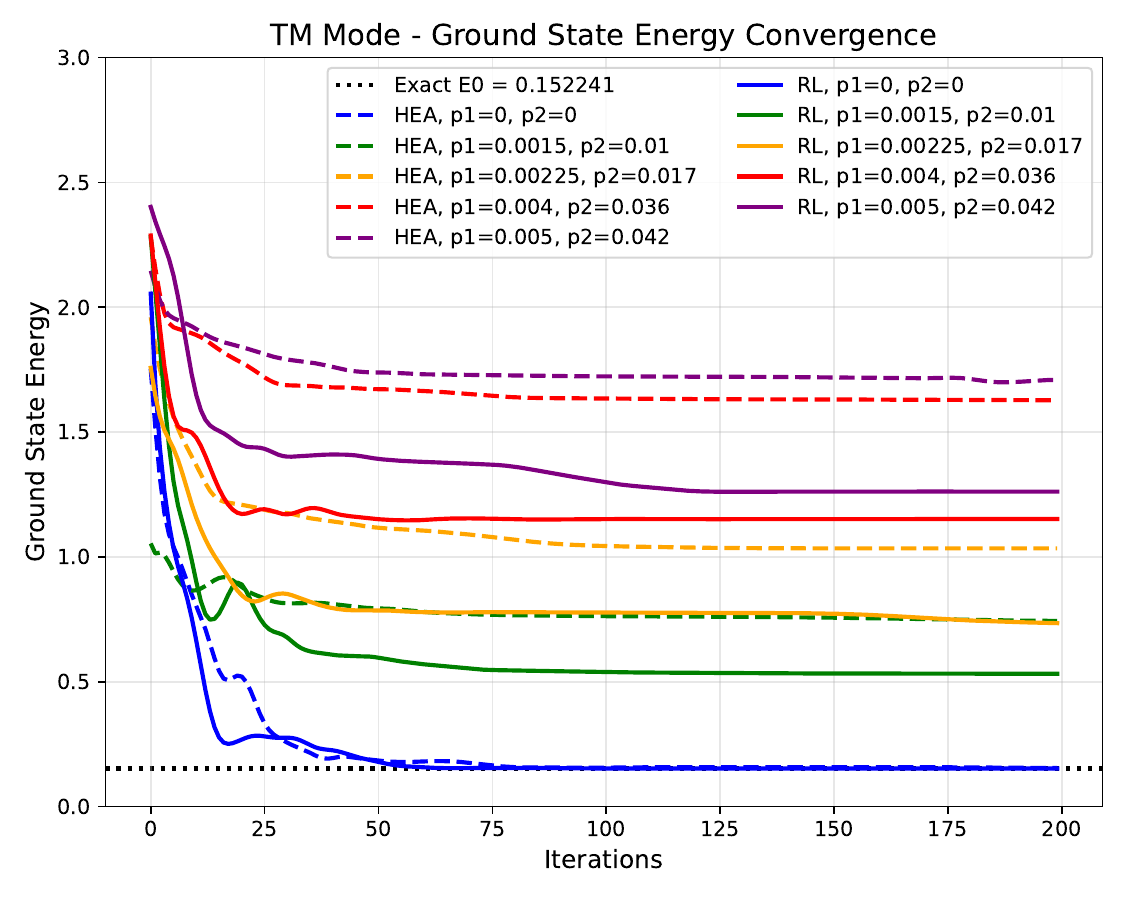} 
    \end{subfigure}
    \begin{subfigure}[b]{0.24\textwidth}
        \centering
        \includegraphics[width=\textwidth]{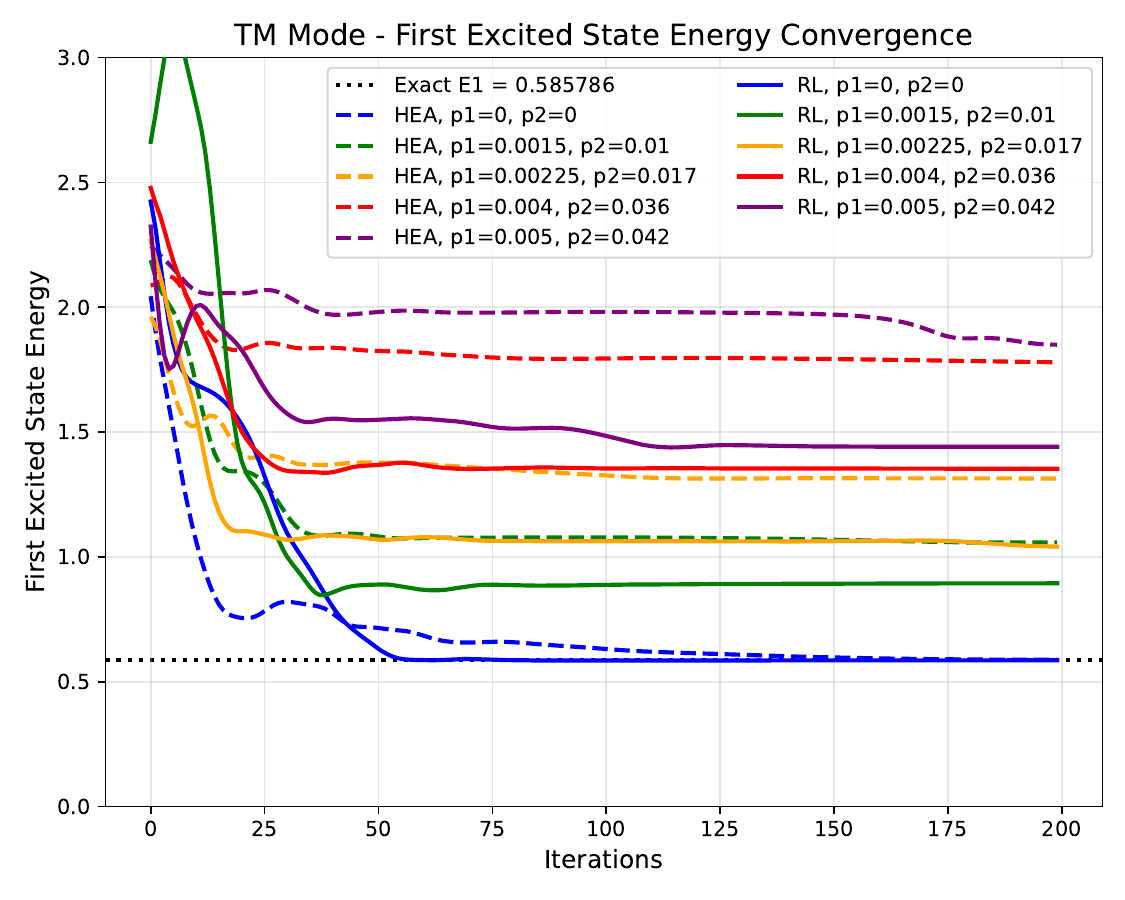} 
    \end{subfigure}
    \begin{subfigure}[b]{0.24\textwidth}
        \centering
        \includegraphics[width=\textwidth]{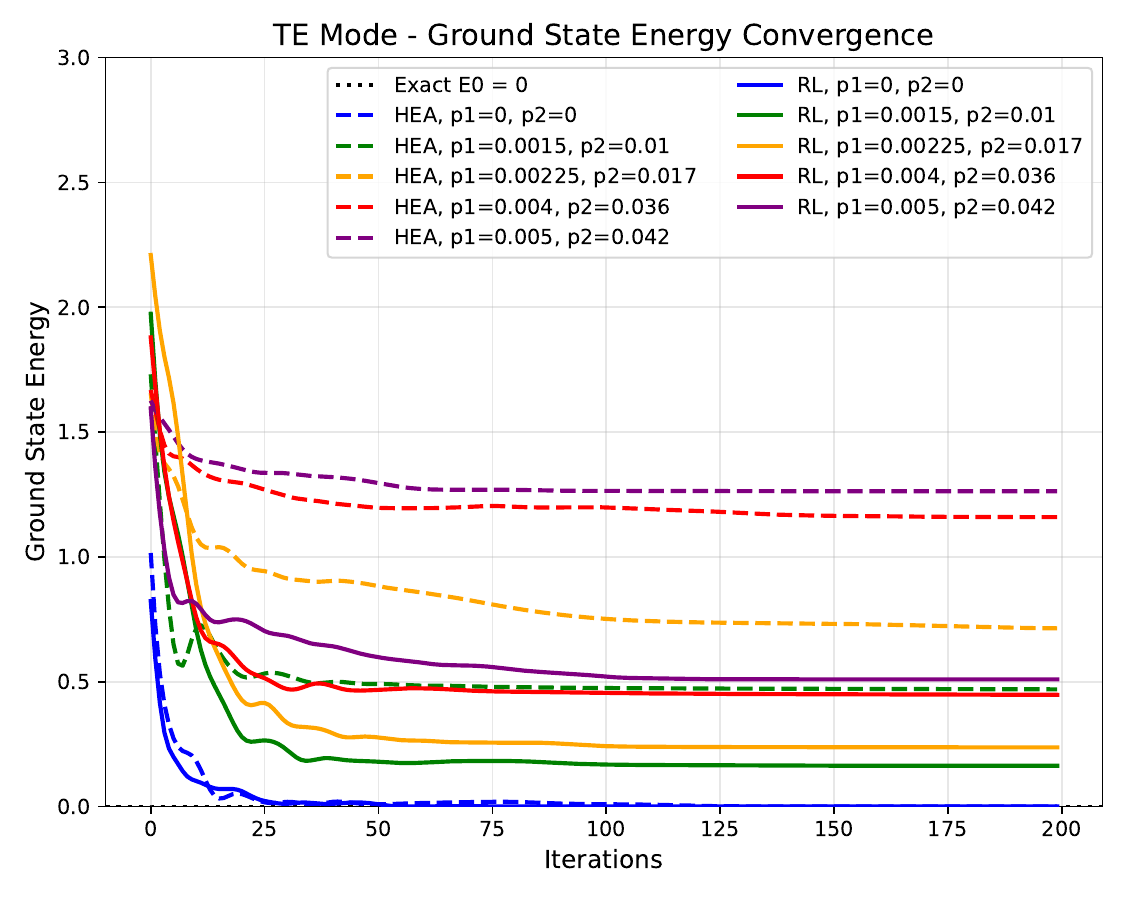}
    \end{subfigure}
    \begin{subfigure}[b]{0.24\textwidth}
        \centering
        \includegraphics[width=\textwidth]{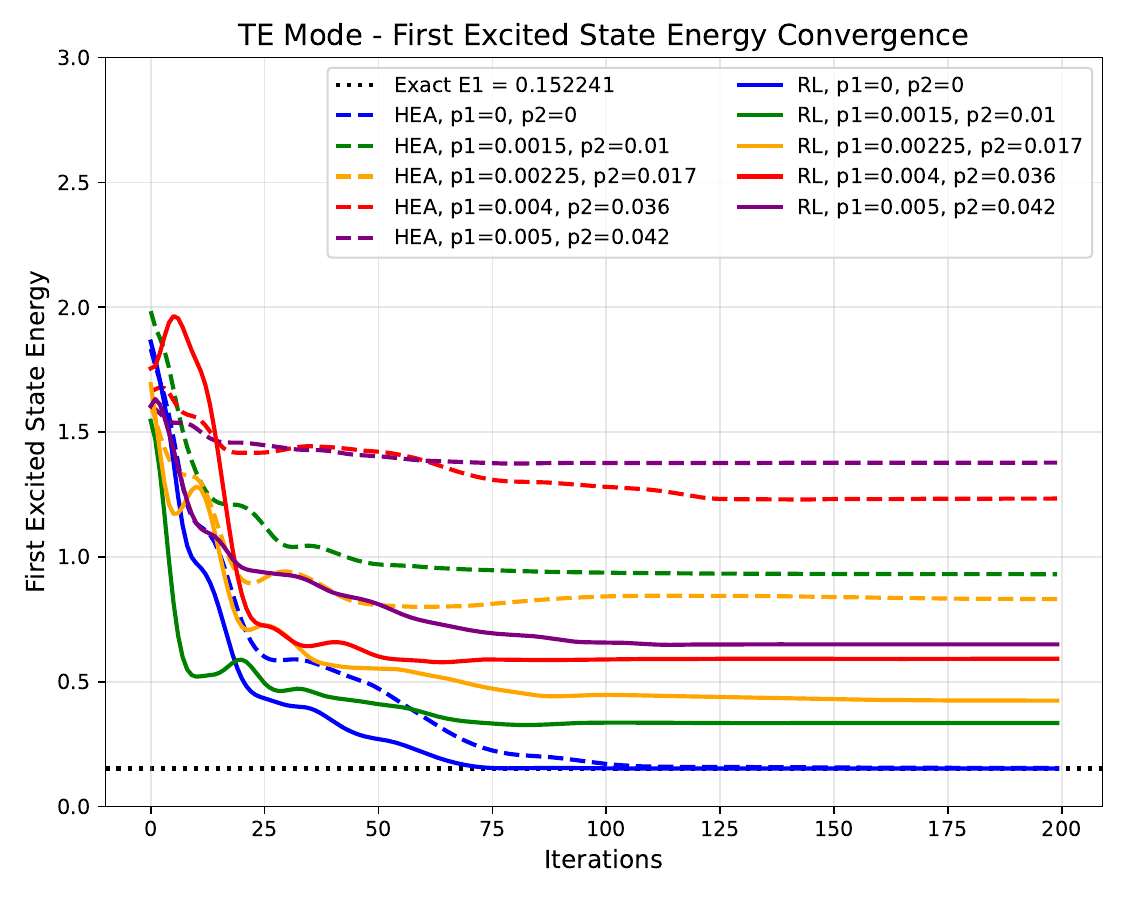} 
    \end{subfigure}
    \caption{Energy convergence comparison of HEA and RL Ansatz under depolarizing noise for TM and TE modes.}
    \label{fig:3-hea-rl-noise}
\end{figure*}
\section{Results}
To assess the performance of our framework on electromagnetic field eigenvalue problems, 
we perform numerical experiments on hollow metallic square waveguides encoded with three and five qubits. The three‑qubit encoding corresponds to an 8~$\times$~8\,mm waveguide and the five‑qubit encoding to a 32~$\times$~32\,mm waveguide, obtained by mapping the cross section onto a grid of $1\,\text{mm}^2$ cells.
Experiments are implemented utilizing the PennyLane quantum computing framework, using the Adam optimizer with a learning rate of 0.1 for parameters update. The RL agent is trained over 500 episodes to explore the circuit design space. To benchmark the efficacy of the proposed framework, the three-qubit and five-qubit systems are respectively compared against HEA baselines of commensurate depth. Performance metrics encompass circuit resource overhead, convergence efficiency, energy estimation accuracy, and noise robustness.

\subsection{Three-Qubit Numerical Experiments}

The proposed framework autonomously generates circuits with reduced gate complexity. As detailed in TABLE~\ref{tab:resource_3qubit}, the total gate count decreases from 30 to 25, accompanied by a decrease in the two-qubit gate proportion from 40\% to 24\%. This significant reduction in entangling operations mitigates the primary source of circuit noise, thereby enhancing fidelity on NISQ hardware.
\begin{table}[htbp]
\centering
\caption{Circuit Resource Comparison for Three-Qubit Systems.}
\begin{tabular}{lccc}
\hline
\textbf{Gate type} & \textbf{6-layer HEA} & \textbf{RL (TM)} & \textbf{RL (TE)} \\
\hline
CNOT gates & 12 & 9 & 6 \\
RY gates & 18 & 16 & 19 \\
\hline
\textbf{Total gates} & 30 & 25 & 25 \\
\textbf{Single-qubit gates} & 18 & 16 & 19 \\
\textbf{Two-qubit gates} & 12 & 9 & 6 \\
\textbf{Single-qubit ratio} & 60\% & 64\% & 76\% \\
\textbf{Two-qubit ratio} & 40\% & 36\% & 24\% \\
\hline
\textbf{Parameters} & 18 & 16 & 19 \\
\textbf{Qubits} & 3 & 3 & 3 \\
\hline
\end{tabular}
\label{tab:resource_3qubit}
\end{table}

\begin{table}[htbp]
\centering
\caption{Eigenvalue Accuracy Comparison for Three-Qubit Waveguide Modes.}
\resizebox{0.5\textwidth}{!}{
\begin{tabular}{llccc}
\hline
\textbf{Mode} & \textbf{Energy metric} & \textbf{6-layer HEA} & \textbf{RL} & \textbf{Analytic} \\
\hline
\multirow{4}{*}{TM} & \(E_0\)  & \(0.152241\) & \(0.152241\) & \(0.152241\) \\
 & \(E_1\) & \(0.585786\) & \(0.585786\) & \(0.585786\) \\
 & \(E_0\) err (\%) & \(0.00000\%\) & \(0.00000\%\) & \(--\) \\
 & \(E_1\) err (\%) & \(0.00000\%\) & \(0.00000\%\) & \(--\) \\
\hline
\multirow{4}{*}{TE} & \(E_0\) & \(3.27974\times 10^{-8}\) & \(3.36930\times 10^{-8}\) & \(0\) \\
 & \(E_1\)  & \(0.152241\) & \(0.152241\) & \(0.152241\) \\
 & \(E_0\) err & \(3.27974\times 10^{-8}\) & \(3.36930\times 10^{-8}\) & \(--\) \\
 & \(E_1\) err (\%) & \(0.00000\%\) & \(0.00000\%\) & \(--\) \\
\hline
\end{tabular}
}
\label{tab:energy_3qubit}
\end{table}

\begin{figure*}[t]
    \centering
    \begin{subfigure}[b]{0.24\textwidth}
        \centering
        \includegraphics[width=\textwidth]{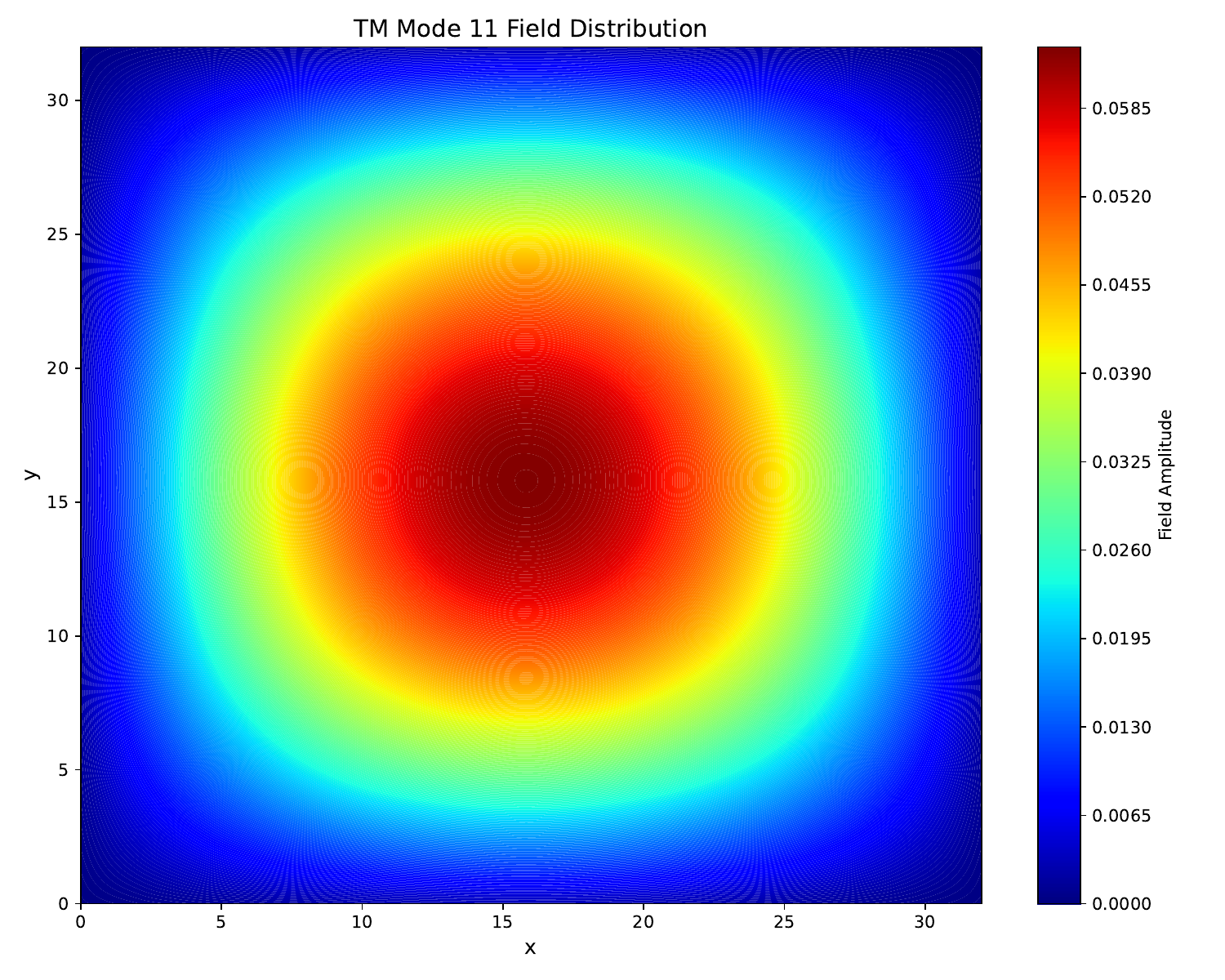} 
    \end{subfigure}
    \begin{subfigure}[b]{0.24\textwidth}
        \centering
        \includegraphics[width=\textwidth]{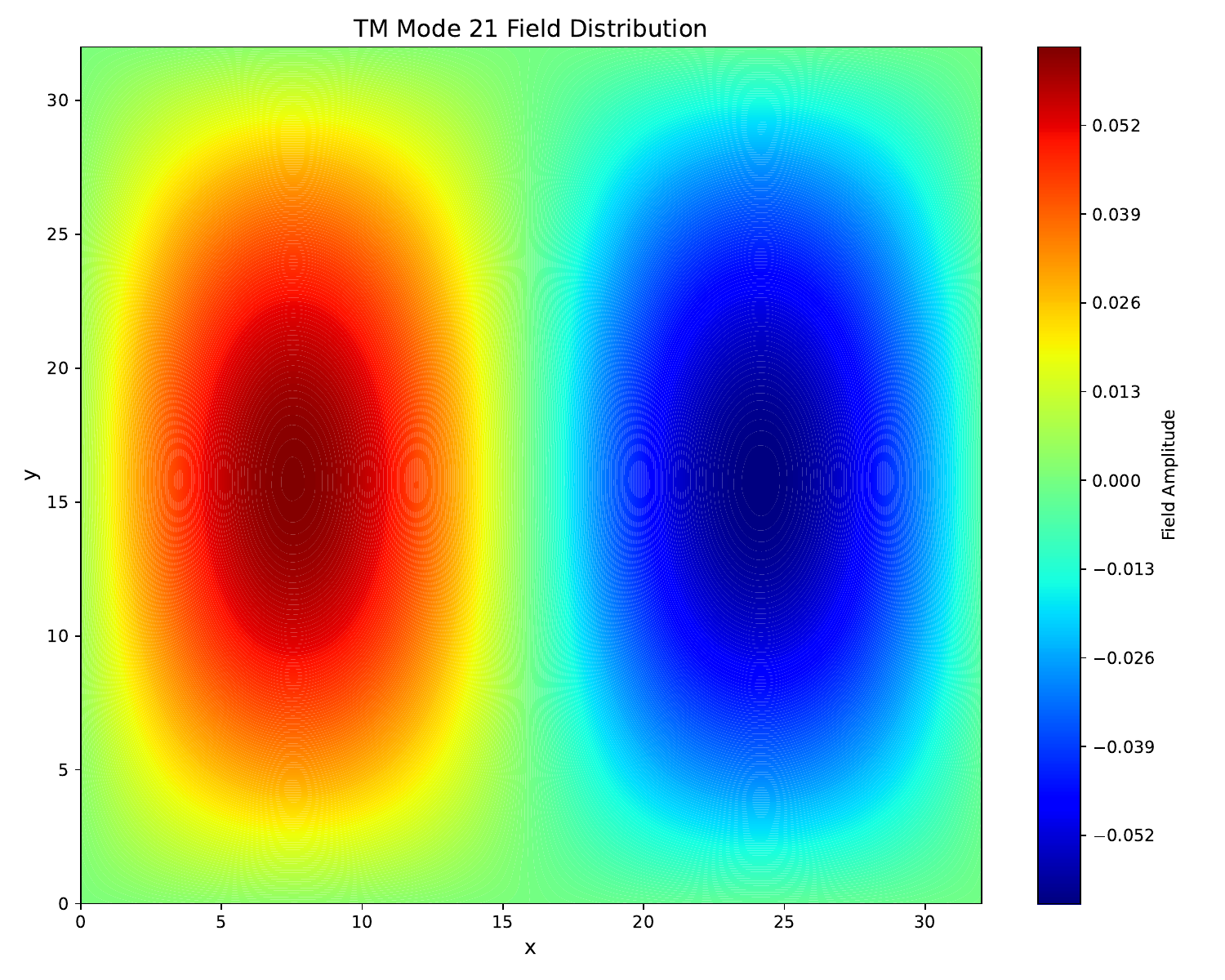} 
    \end{subfigure}
    \begin{subfigure}[b]{0.24\textwidth}
        \centering
        \includegraphics[width=\textwidth]{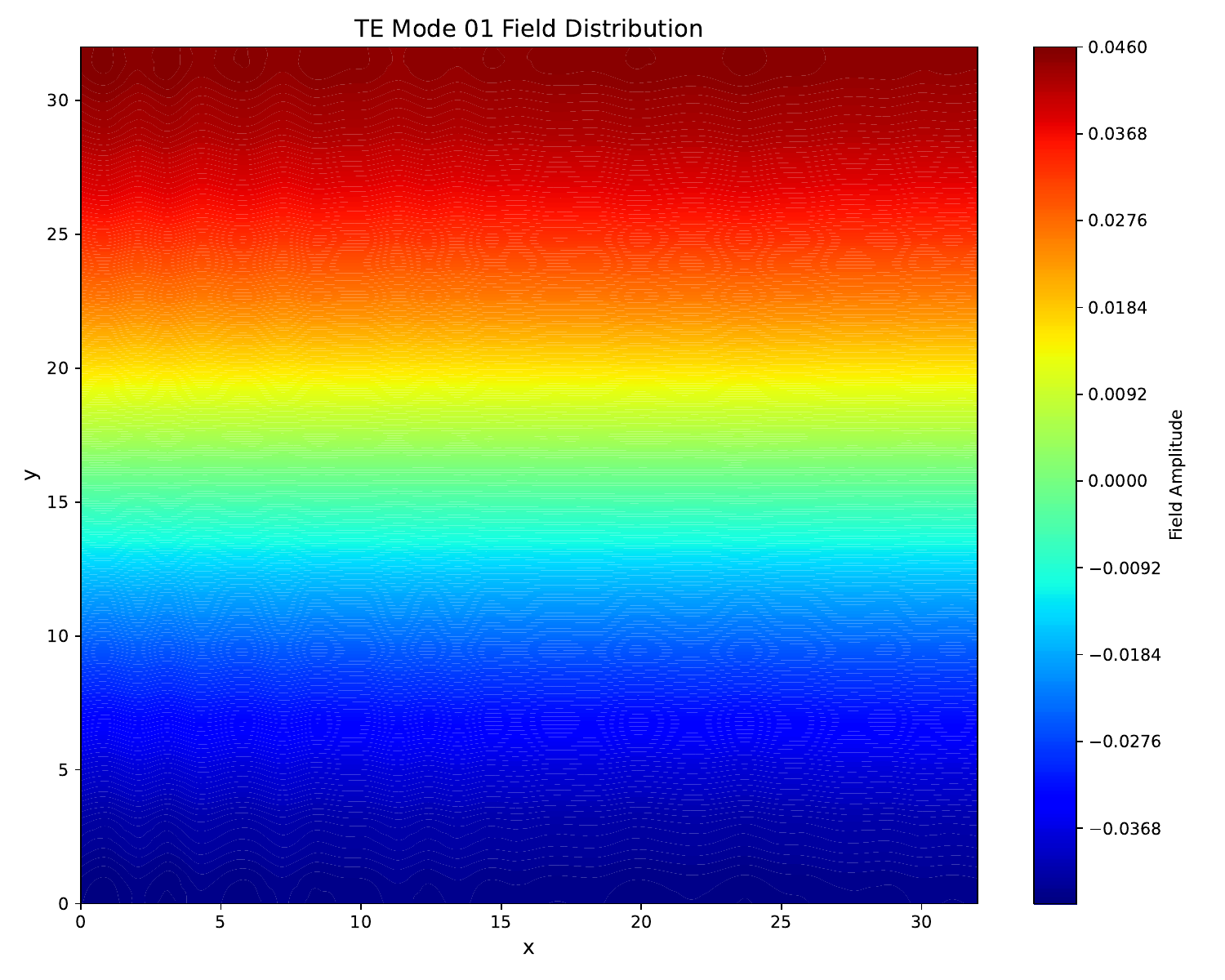}
    \end{subfigure}
    \begin{subfigure}[b]{0.24\textwidth}
        \centering
        \includegraphics[width=\textwidth]{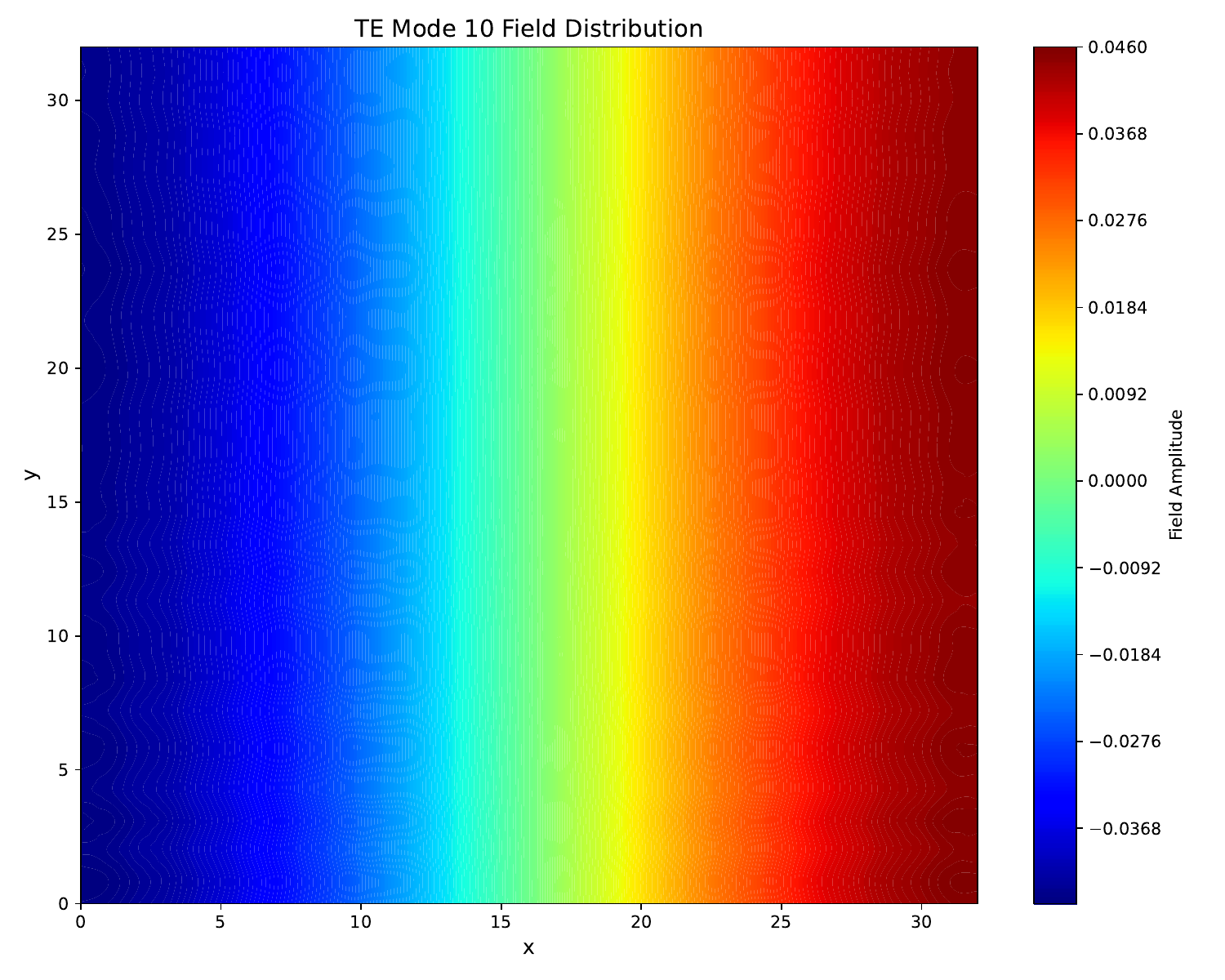} 
    \end{subfigure}
    \caption{Reconstructed field distributions for five-qubit waveguide modes from the obtained eigenstates.}
    \label{fig:field_ideal_5q}
\end{figure*}
\begin{figure*}[t]
    \centering
    \begin{subfigure}[b]{0.24\textwidth}
        \centering
        \includegraphics[width=\textwidth]{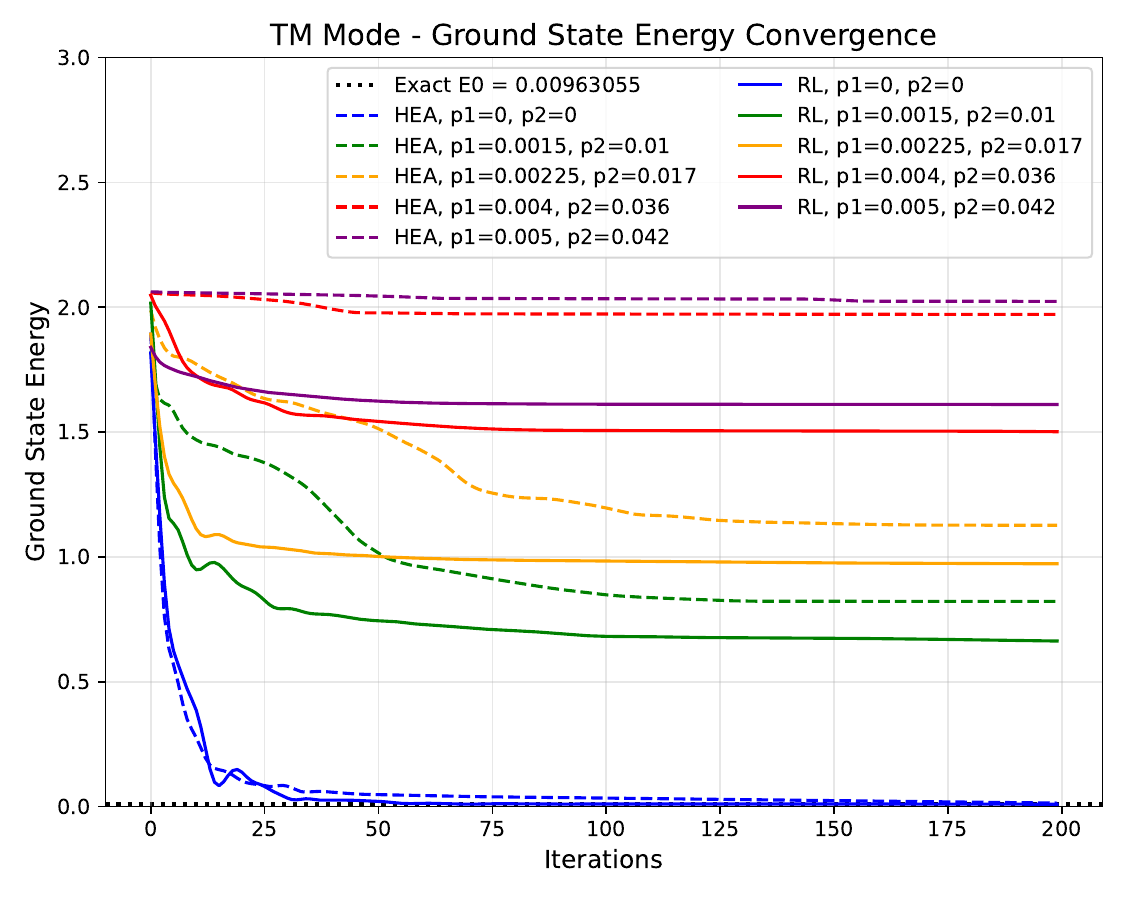} 
    \end{subfigure}
    \begin{subfigure}[b]{0.24\textwidth}
        \centering
        \includegraphics[width=\textwidth]{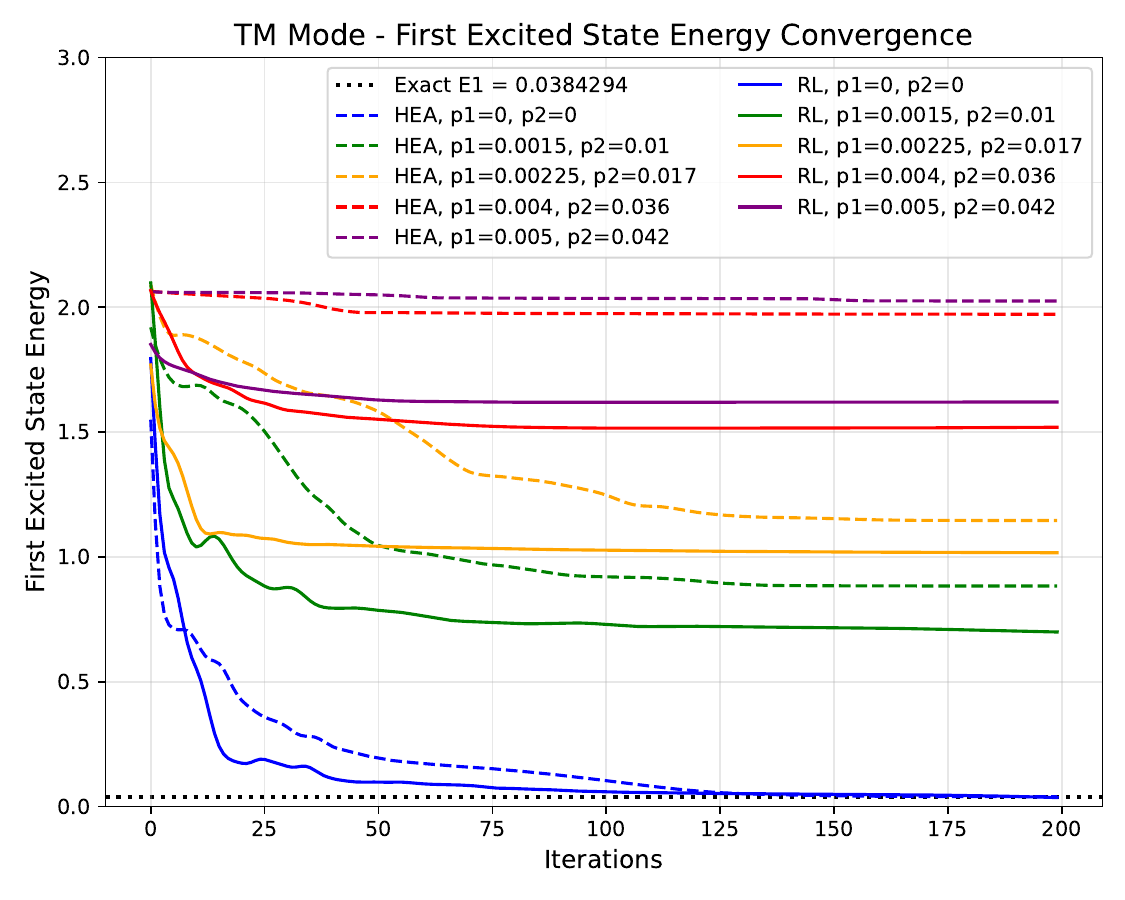} 
    \end{subfigure}
    \begin{subfigure}[b]{0.24\textwidth}
        \centering
        \includegraphics[width=\textwidth]{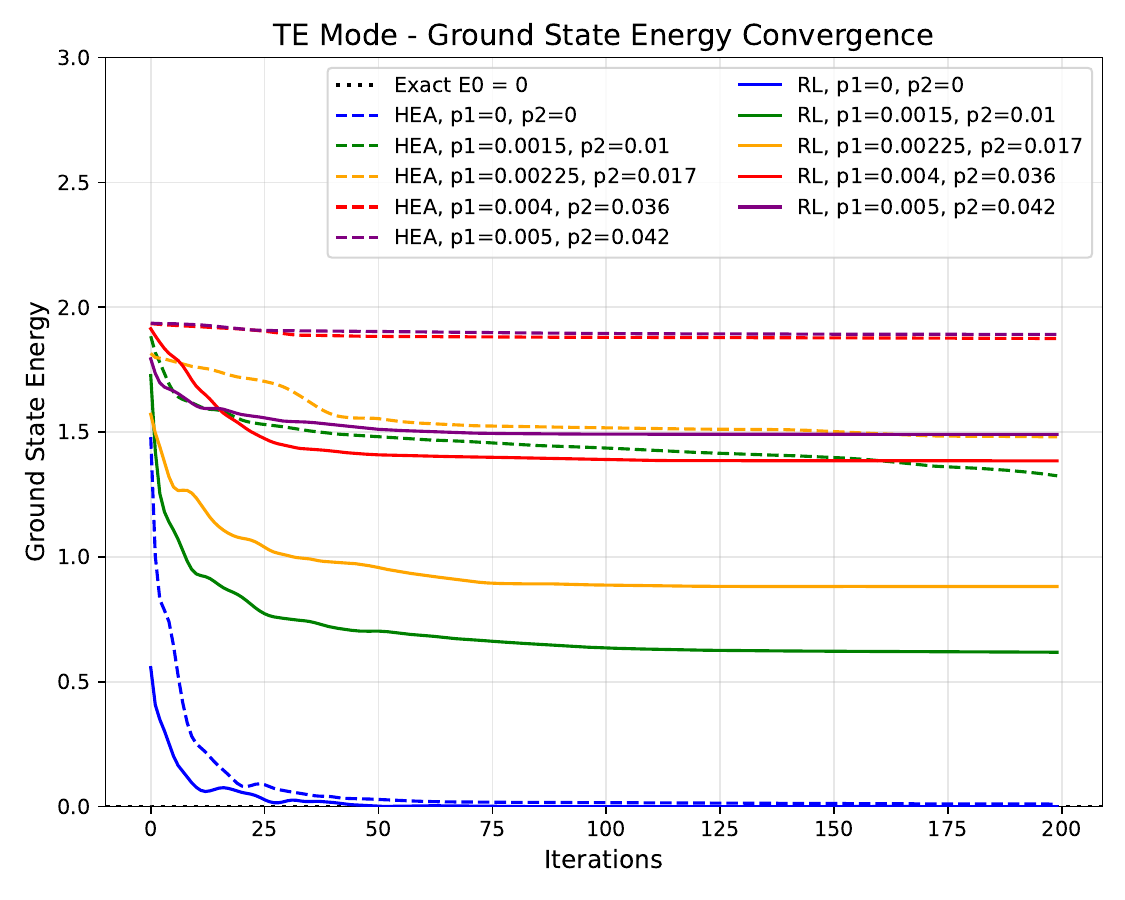}
    \end{subfigure}
    \begin{subfigure}[b]{0.24\textwidth}
        \centering
        \includegraphics[width=\textwidth]{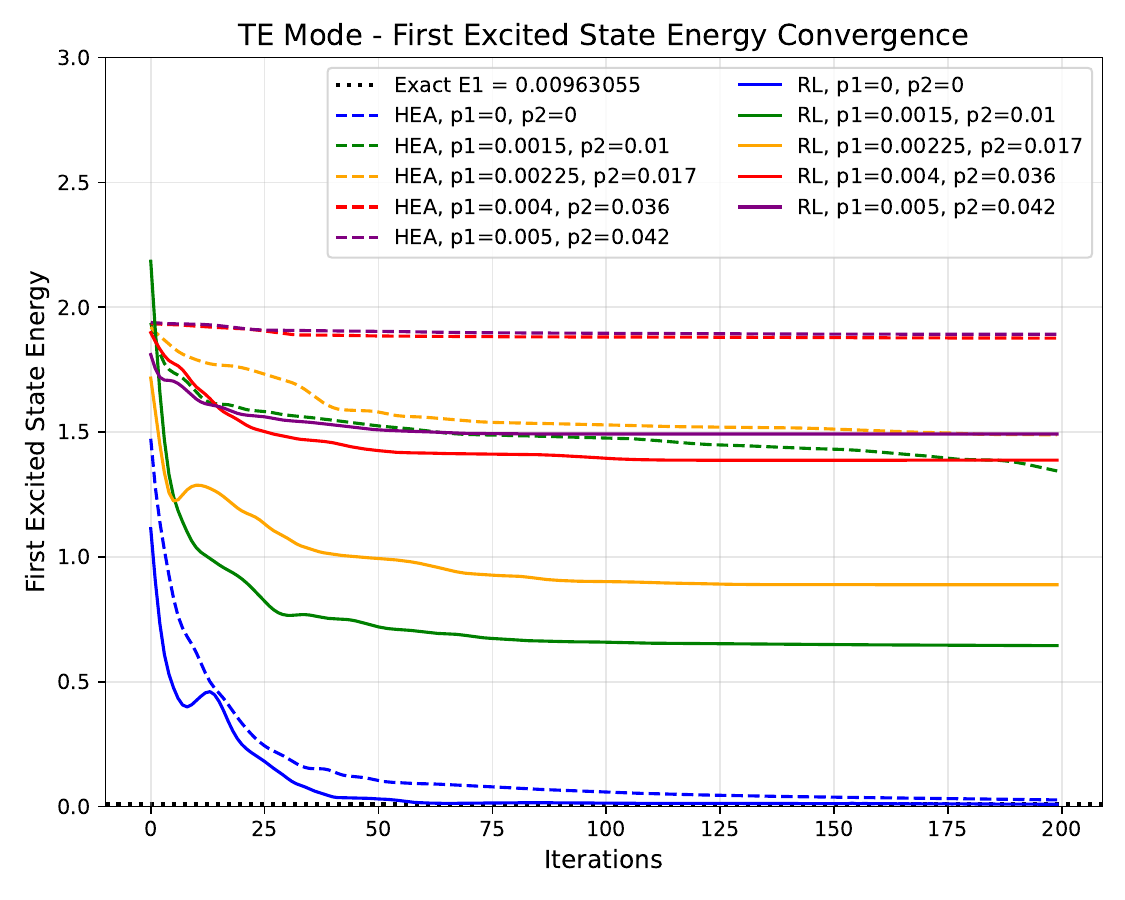} 
    \end{subfigure}
    \caption{Energy convergence comparison of HEA and RL Ansatz under depolarizing noise for TM and TE modes.}
    \label{fig:5-hea-rl-noise}
\end{figure*}

\begin{figure}[htbp]
    \centering 
    \begin{subfigure}[b]{0.48\textwidth}
        \centering
        \includegraphics[width=\textwidth]{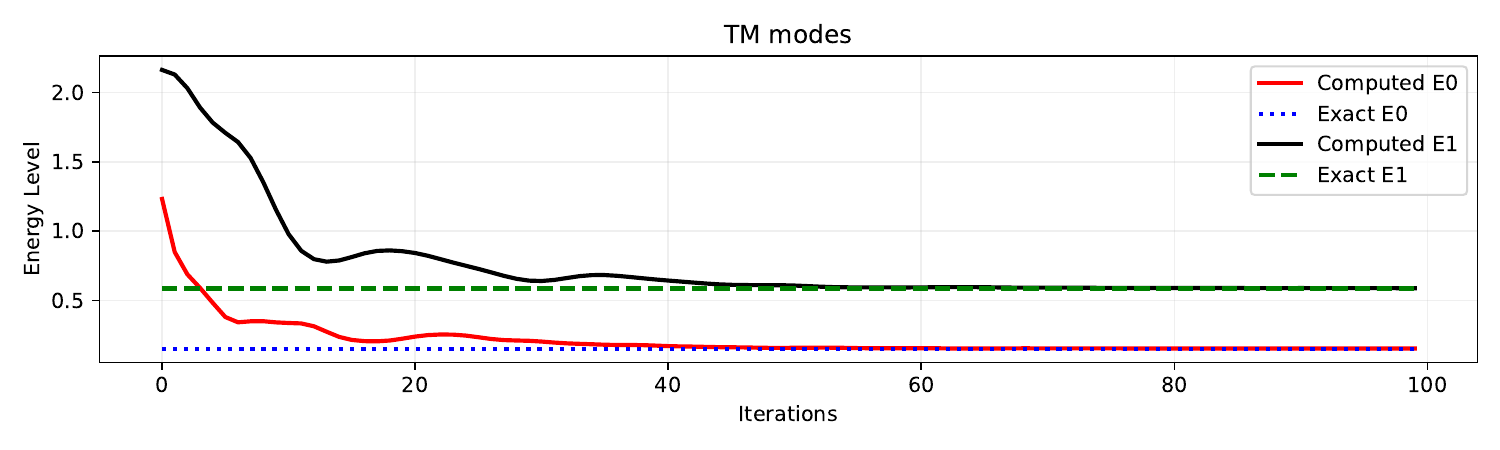}
        \label{fig:convergence_tm}
    \end{subfigure}
    \begin{subfigure}[b]{0.48\textwidth}
        \centering
        \includegraphics[width=\textwidth]{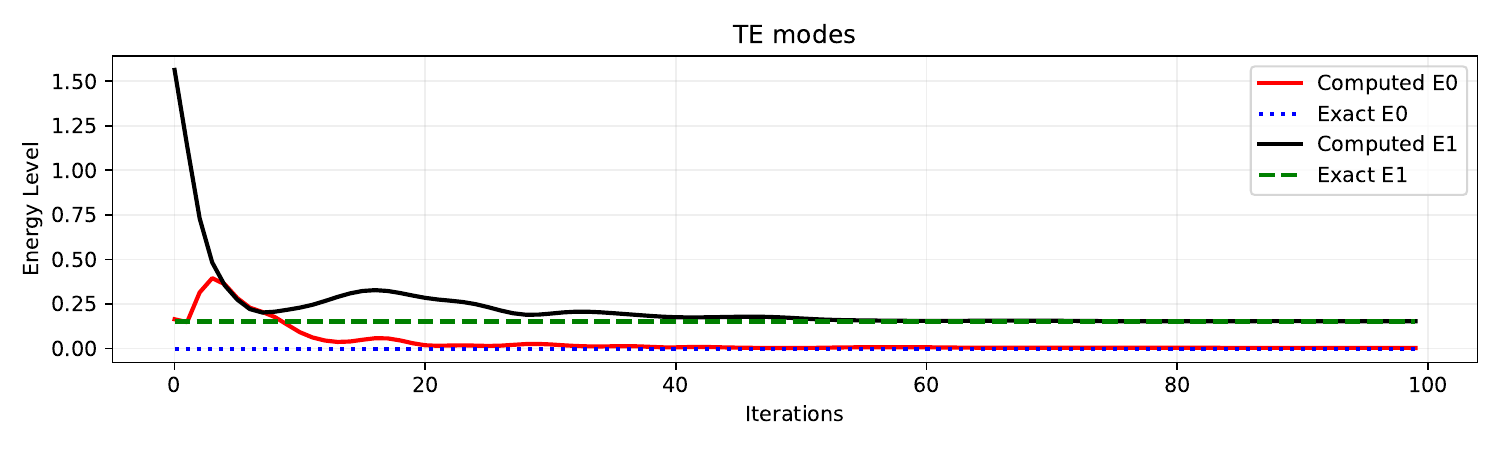}
        \label{fig:convergence_te}
    \end{subfigure}
    \caption{Energy convergence for three-qubit waveguide modes under ideal conditions.}
    \label{fig:convergence_3q}
\end{figure}

Fig.~\ref{fig:convergence_3q} depicts the convergence behavior of the algorithm under ideal noise-free conditions. For both TM and TE modes, the RL-generated circuits attain fast convergence within approximately 50 iterations, showing over a 20-fold acceleration relative to conventional VQE approaches\cite{colella2023shot}. The energy estimation accuracy is quantified in TABLE~\ref{tab:energy_3qubit}, demonstrating that our work and HEA obtain similar or comparable accuracy, matching analytical solutions exactly. For the TE$_{01}$ mode, the ground-state energy deviation remains on the order of $10^{-8}$, confirming the high-fidelity characteristics of the proposed method.

Fig.~\ref{fig:field_ideal_3q} presents the reconstructed transverse electromagnetic field profiles for the TM$_{11}$, TM$_{21}$, TE$_{01}$, and TE$_{10}$ modes under ideal noise-free situations. The spatial patterns exhibit strong agreement with classical electromagnetic predictions, validating the correct mapping from quantum states to physical field configurations. To evaluate robustness under realistic quantum hardware, we incorporate depolarizing noise models across four distinct noise strengths. 
Fig.~\ref{fig:3-hea-rl-noise} present energy convergence trajectories for both RL-generated circuits and the HEA baseline. The circuits generated by RL consistently maintain smaller estimation errors at all tested noise levels.
This enhanced resilience is attributable to the shallower circuit depth, which effectively limits stochastic gate error accumulation. Fig.~\ref{fig:field_noise_3q} visualizes the reconstructed field distributions under noisy conditions. Compared to the ideal case, depolarizing noise introduces conspicuous spatial distortions and amplitude fluctuations, underscoring its severe impact on field reconstruction fidelity.

\subsection{Five-Qubit Numerical Experiments}

\begin{figure}[htbp]
    \centering 
    \begin{subfigure}[b]{0.48\textwidth}
        \centering
        \includegraphics[width=\textwidth]{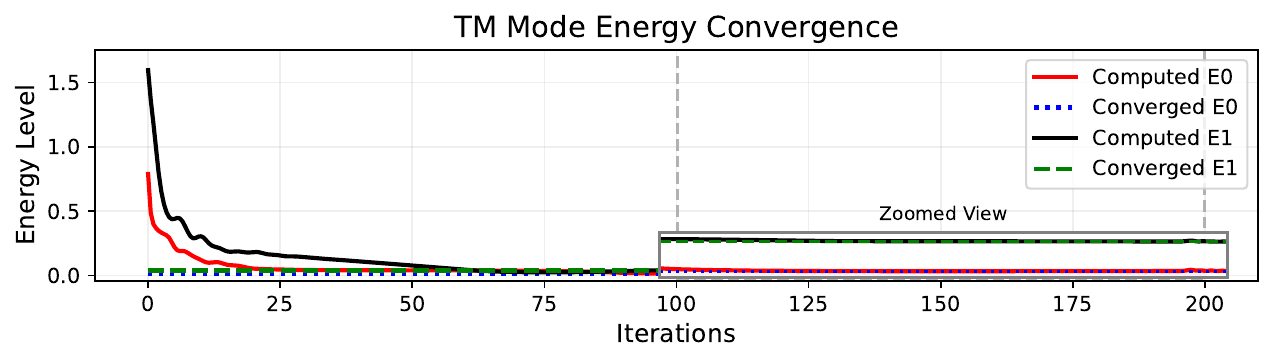}
        \label{fig:convergence_tm}
    \end{subfigure}
    \begin{subfigure}[b]{0.48\textwidth}
        \centering
        \includegraphics[width=\textwidth]{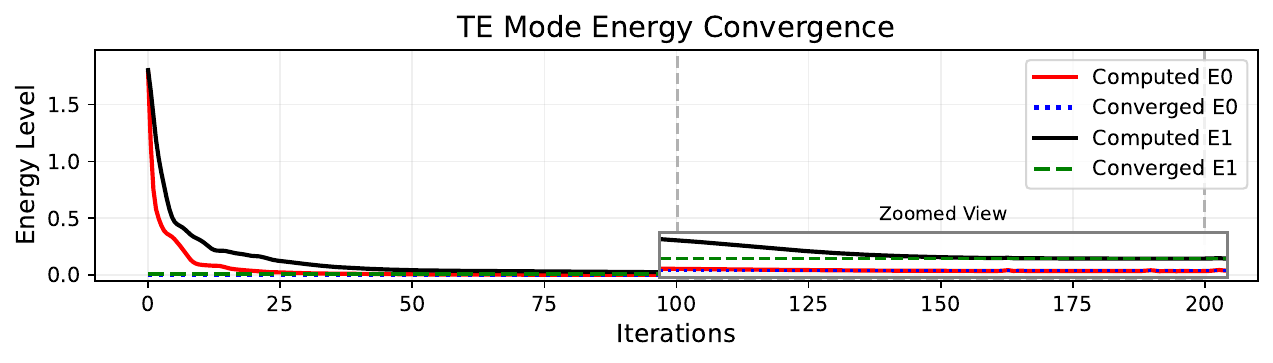}
        \label{fig:convergence_te}
    \end{subfigure}
    \caption{Energy convergence for five-qubit waveguide modes under ideal conditions.}
    \label{fig:convergence_5q}
\end{figure}

Next, extending from three to five qubits increases the Hamiltonian representation to 47 Pauli operators (Eqs.~\eqref{13}--\eqref{14}), which allows for more refined characterization of electromagnetic field distributions in rectangular waveguides.
TABLE~\ref{tab:resource_5qubit} presents a quantitative comparison of circuit resources in terms of gate count and circuit depth.
The RL framework autonomously constructs circuits with total gate counts of 90--100 versus 135 for the 15-layer HEA, and trainable parameters numbering 47--57 versus 75, demonstrating amplified resource savings at larger scales.
As illustrated in Fig.~\ref{fig:convergence_5q} and TABLE~\ref{tab:energy_5qubit}, RL-based SSVQE reaches target accuracy within fewer than 150 iterations with energy relative errors below 2\% for all modes. These results confirm that the adaptive architecture strategy successfully preserves computational accuracy as the system size increases.
Fig.~\ref{fig:field_ideal_5q} displays the constructed
transverse electromagnetic field profile under ideal conditions for the five-qubit system, revealing finer spatial details  compared to their three-qubit counterparts.

\begin{table}[!htbp]
\centering
\caption{Circuit Resource Comparison for Five-Qubit Systems.}
\begin{tabular}{lccc}
\hline
\textbf{Gate type} & \textbf{15-layer HEA} & \textbf{RL (TM)} & \textbf{RL (TE)} \\
\hline
CNOT gates & 60 & 43 & 42 \\
RY gates & 75 & 47 & 57 \\
\hline
\textbf{Total gates} & 135 & 90 & 99 \\
\textbf{Single-qubit gates} & 75 & 47 & 57 \\
\textbf{Two-qubit gates} & 60 & 43 & 42 \\
\textbf{Single-qubit ratio} & 55.6\% & 52.2\% & 57.6\% \\
\textbf{Two-qubit ratio} & 44.4\% & 47.8\% & 42.4\% \\
\hline
\textbf{Parameters} & 75 & 47 & 57 \\
\textbf{Qubits} & 5 & 5 & 5 \\
\hline
\end{tabular}
\label{tab:resource_5qubit}
\end{table}

\begin{table}[!htbp]
\centering
\caption{Eigenvalue Accuracy Comparison for Five-Qubit Waveguide Modes.}
\resizebox{0.5\textwidth}{!}{
\begin{tabular}{llccc}
\hline
\textbf{Mode} & \textbf{Energy metric} & \textbf{15-layer HEA} & \textbf{RL} & \textbf{Analytic} \\
\hline
\multirow{4}{*}{TM} & \(E_0\)  & \(0.00966393\) & \(0.00969766\) & \(0.00963055\) \\
 & \(E_1\) & \(0.0384386\) & \(0.0386376\) & \(0.0384294\) \\
 & \(E_0\) err (\%) & \(0.346600\%\) & \(0.696800\%\) & \(--\) \\
 & \(E_1\) err (\%) & \(0.0239400\%\) & \(0.541700\%\) & \(--\) \\
\hline
\multirow{4}{*}{TE} & \(E_0\) & \(0.000111694\) & \(0.0000363846\) & \(0\) \\
 & \(E_1\)  & \(0.00972522\) & \(0.00978106\) & \(0.00963055\) \\
 & \(E_0\) err & \(1.11694 \times 10^{-4}\) & \(3.63846 \times 10^{-5}\) & \(--\) \\
 & \(E_1\) err (\%) & \(0.982900\%\) & \(1.56300\%\) & \(--\) \\
\hline
\end{tabular}
}
\label{tab:energy_5qubit}
\end{table}

Fig.~\ref{fig:5-hea-rl-noise} illustrates convergence curves under four noise intensities, confirming the advantage of RL circuits over HEA in maintaining convergence robustness under noisy conditions.

Collectively, the experimental results establish the efficacy of the proposed framework across multiple system scales: the adaptive architecture exploration strategy achieves gate-count reduction, convergence acceleration, and enhanced noise tolerance relative to fixed-architecture baselines in both three-qubit and five-qubit configurations. The high-fidelity field reconstructions under ideal conditions validate algorithmic correctness, while the pronounced field degradation observed in noisy experiments highlights the critical importance of noise mitigation in NISQ devices. The five-qubit results further confirm the framework's scalability, as the adaptive strategy retains resource optimization advantages even when confronting higher-dimensional Hamiltonians.

\section{Conclusion}

This work has presented an architecture and shot adaptive SSVQE that addresses the resource inefficiency and noise vulnerability inherent in conventional VQE-based electromagnetic eigenmode solvers. By integrating reinforcement learning-driven automated circuit synthesis with adaptive shot allocation, the proposed framework reduces redundant gate structures and optimizes measurement resource distribution.
Numerical experiments on three-qubit and five-qubit rectangular waveguide systems demonstrate gate-count reductions of up to 45 gates, convergence acceleration exceeding 20-fold, and energy estimation accuracies consistent with analytical solutions. 

The adaptive architecture strategy effectively accommodates the hardware constraints of NISQ devices, establishing a practical quantum algorithmic pathway for electromagnetic eigenmode analysis in microwave engineering applications. 
In future work, we intend to extend the framework's applicability from rectangular waveguides to cylindrical, elliptical, and arbitrary cross-sectional geometries. Additionally, we aim to leverage higher qubit counts to achieve finer-grained simulations of these complex structures.
As quantum hardware continues to mature, RL-based SSVQE holds promise for practical deployment in the computational design of complex microwave components and systems.

    \section*{Acknowledgment}
    National Natural Science Foundation of China (Grant No. 62471126), the Jiangsu Frontier Technology Research and Development Plan (Grant No. BF2025066), and the Fundamental Research Funds for the Central Universities (Grant No. 2242022k60001).


    %





    \ifCLASSOPTIONcaptionsoff
    \newpage
    \fi





    \bibliographystyle{IEEEtran}
    \bibliography{IEEEabrv,Bibliography}

    \vfill


\end{document}